# Generative vector search to improve pathology foundation models across multimodal vision-language tasks

Markus Ekvall[1*], Ludvig Bergenstråhle[1], Patrick Truong[1], Ben Murrell[2], and Joakim Lundeberg[1*]

[1]Science for Life Laboratory, School of Engineering Sciences in Chemistry, Biotechnology, and Health, Royal Institute of Technology – KTH, Box 1031, SE-17121 Solna, Sweden
[2]Department of Microbiology, Tumor and Cell Biology, Karolinska Institutet, SE-17177 Solna, Sweden
[*]Corresponding authors, markus.ekvall@scilifelab.se and joakim.lundeberg@scilifelab.se

**Abstract**

Retrieval-augmented generation improves large language models by grounding outputs in external knowledge sources, reducing hallucinations and addressing knowledge cutoffs. However, standard embedding-based retrieval fails to capture the complexity of multi-concept queries, particularly in domains like biomedicine, where biological data are inherently high-dimensional. For example,omics datasets, and clinical reports simultaneously exhibit numerous molecular, cellular, and physiological features. We present **St**o**ch**astic **L**atent **M**atching (**STHLM**), a generative vector search method that samples query-conditioned embeddings from text or image inputs to enhance retrieval performance. Analogous to how Chain-of-Thought reasoning enables language models to "think longer" on complex problems, STHLM allows retrieval systems to "search wider" through iterative sampling. STHLM demonstrates critical improvements over classical vector retrieval across diverse benchmarks, including scientific literature, clinical notes, and tissue images, boosting retrieval performance by 10-30% through test-time compute (trading latency for accuracy), while enabling up to a 10-fold compression of embedding dimensions.

**Main**

Large language models (LLMs) show potential for diverse biomedical applications, including clinical documentation, diagnostic support, and multimodal integration of imaging with clinical data for treatment planning[1]. Recent advances in test-time compute, such as Chain-of-Thought (CoT) reasoning[2], enable LLMs to "think longer" on complex problems, trading latency for improved accuracy. However, knowledge cutoffs, training biases, and generation of unverifiable outputs pose significant challenges for deployment in biomedical domains. These limitations can result in responses that omit recent findings or provide incorrect guidance with high confidence[3] – critical issues in precision-sensitive fields such as biomedicine[3]. Retrieval-augmented generation (RAG) addresses these challenges by grounding model outputs in verified, up-to-date knowledge sources. However, the retrieval systems that form the critical first step in RAG pipelines lack an analogous mechanism to CoT to "search wider" when faced with challenging multi-concept queries.

RAG systems consist of a data source, a retriever, and an LLM. Given a query, the LLM prompts the retriever to fetch relevant documents that can be included in the context of the LLM to ground the response. Although RAG frameworks help mitigate several limitations of LLMs, they introduce challenges of their own. Retrieval accuracy remains the central challenge, as even slight retrieval errors can substantially reduce performance despite the availability of relevant data[3]. To address this, major AI companies (e.g., Google[4], OpenAI[5], Alibaba[6], NVIDIA[7]) have developed large-scale general-purpose encoders trained on massive datasets and compute. Meanwhile, academia has focused on domain-specific models for pathology[8–10], dermatology[11], single-cell biology[12], as well as transcriptomics, genomics, and proteomics[13].

Retrieval in RAG systems typically utilizes a vector database populated with encoder-generated embeddings (Fig. 1a), where an encoder assigns a semantic position to each entry. Queries are encoded to match the most relevant entries (Fig. 1b) for retrieval. However, vector databases face significant challenges: high-dimensional embeddings are required to capture complex data, but this increases training and memory costs, and queries often span multiple concepts occupying different regions of embedding space. These retrieval difficulties arise from a fundamental capacity constraint: when forced to represent diverse, multi-concept data in limited embedding dimensions, the system struggles to maintain clear separation between different semantic categories. These limitations have proven so significant that Google DeepMind is advocating for new research directions[14]. We demonstrate how generative modelling offers a natural solution to the limitations of embedding-based retrieval on a toy dataset in the supplementary section *Vector Database Capacity Limitations and Generative Modelling.*

To verify generative modelling as a solution on real-world data, we introduce Stochastic Latent Matching (STHLM), a generative vector search framework inspired by generative models like DALL-E 2[15] and Stable Diffusion[16], which condition on text embeddings to generate images. STHLM trains a conditional flow-matching model (although any generative model would work) that takes a query embedding (from text or images) and transforms random Gaussian noise into multiple embeddings that represent the semantic distribution of the query (Fig. 1c). Unlike conventional methods that produce a single embedding (Supplementary Fig. 2a), STHLM models the full conditional distribution, capturing the multimodal nature of complex queries such as "tissue showing inflammation and necrosis."

By generating *N* samples from this distribution, users can trade latency for improved retrieval accuracy—analogous to how CoT reasoning allows language models to "think longer" on complex problems (Fig. 1d–f). This sampling-based approach is particularly valuable for multi-concept queries where a single point estimate may lie closer to irrelevant documents than to the diverse set of relevant ones (Supplementary Fig. 2a–b).

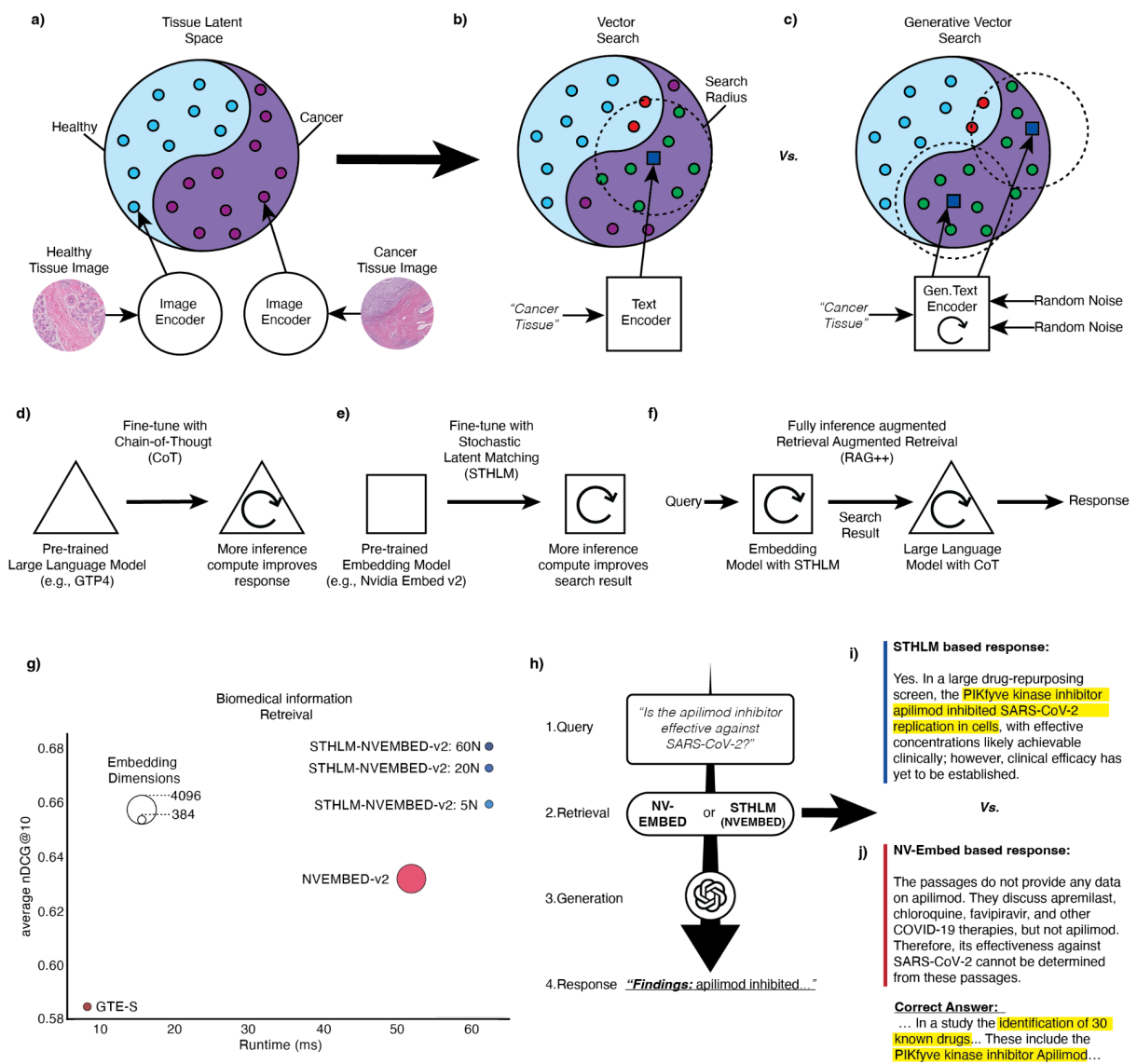

**Figure 1: Illustrations and benchmarks demonstrating how STHLM enables embedding models to use inference compute for a "wider search" to improve search results.**

**a)** A tissue image encoder maps samples into latent space: blue points = healthy tissue; purple = cancerous tissue.

**b)** In conventional vector search, a query like "Cancer Tissue" is encoded as a point (blue square); retrieval is based on proximity (dashed circle). Green = correct; red = incorrect results.

**c)** In generative vector search, a conditional model generates a distribution of embeddings for a query like "Cancer Tissue" instead of a single point. Random noise drawn from a Gaussian is transformed into query-conditioned samples (blue squares) using a conditional generative model. Images are retrieved based on proximity to these samples (dashed circles), yielding more relevant results compared to incorrect ones, as seen in conventional search.

**d)** LLMs are currently trained in two steps: unsupervised pre-training and then fine-tuned to think longer with Chain-of-Thought. The latter step allows the LLM to think for longer on more complex tasks to give a more accurate response.

**e)** Currently embedding models are only trained with unsupervised pre-training. However, with STHLM we have a new fine-tuning technique to allow the embedding model to use a wider search to find more relevant results.

**f)** With STHLM, RAG systems can increase inference compute in the two major steps (search and respond), allowing the system to search wider and think longer on more complex tasks.

**g)** Average retrieval performance comparison across three biomedical benchmarks (SciFact, NFCorpus, and BioASQ, see Supplementary Fig. 3 for individual benchmark results). Y-axis: Average performance measured by Normalized Discounted Cumulative Gain at rank 10

(NDCG@10, higher is better). X-axis: Inference time in milliseconds (lower is better). Point size: Embedding dimensionality (smaller points = fewer dimensions = lower memory cost). STHLM-NV-Embed-v2 generates target embeddings using an Euler ODE solver with 4 steps; increasing sample counts (N) requires no additional time since samples are generated in parallel. See Supplementary Fig. 6 for the effect of ODE solver steps on the performance-latency trade-off.

**h)** Overview of RAG. The user provides a query (1), which is used by the retrieval system (2), either NV-Embed-v2 or STHLM-NV-Embed-v2, to retrieve relevant clinical records. The LLM (3) is then prompted with both the retrieved clinical records and the original query to generate a response (4).

**i–j)** A representative example from MultiCare shows retrieval using a synthetic query with STHLM-NV-Embed-v2 (**i**) and NV-Embed-v2 (**j**). Both systems use the same query and database, with retrieved records passed to ChatGPT for response generation. NV-Embed-v2 returns peripherally relevant records, but no records answering the query. In contrast, STHLM-NV-Embed-v2 retrieves more contextually relevant records, enabling the chatbot to answer the question.

We evaluated STHLM on three distinct text benchmarks relevant for biomedical RAG. First, on SciFact[17] (text-to-text scientific retrieval), second on NFCorpus[18] (biomedical information retrieval), and third on BioASQ[19] (biomedical Q&A). We compared STHLM with General-purpose Text Embedding model Small (GTE-S, 30M parameters, 384 embedding dimensions)[6], which is one of the best small text encoders, and the current state-of-the-art NV-Embed-v2[20] (approximately 7B parameters, 4096 embedding dimensions). We trained STHLM on dimensionality-reduced embeddings (384 dimensions) of NV-Embed-v2 using

Principal Component Analysis (PCA), yielding STHLM-NV-Embed-v2, to decrease memory and training requirements, but still have enough dimensions to make STHLM effective. As shown in Fig. 1g (see Supplementary Fig. 3 for individual benchmark results), by adding approximately 20% more inference compute to NV-Embed-v2, STHLM boosted its retrieval performance by an amount comparable to the improvement NV-Embed-v2 provides over the much smaller GTE-S model, while STHLM requires 10-fold fewer embedding dimensions. To illustrate the potential impact in medical applications, we compared a RAG system using either STHLM-NV-Embed-v2 or NV-Embed-v2 as the retriever (Fig. 1h–j), showing that the former enabled the language model to produce more helpful outputs.

We further evaluated STHLM in pathology by benchmarking it against three state-of-the-art pathology image-caption foundation models: Pathology Language and Image Pre-training (PLIP)[8], QuiltNet[9], and PathGen-CLIP[10]. While all follow the Contrastive Language–Image Pretraining (CLIP)[21] approach, they differ mainly in training scale: PLIP used 300,000 examples, QuiltNet 1 million, and PathGen-CLIP 2 million (1.6 million from PathGen1.6M plus other sources). We trained STHLM to align text and image embeddings from PathGen-CLIP, producing STHLM-PathGen. On zero-shot classification tasks, despite training on only 307,000 examples, STHLM-PathGen consistently outperformed PathGen-CLIP across datasets (Fig. 2a). We also evaluated it in multi-label classification with four labels[22] (Supplementary Fig. 4c). As shown in Supplementary Fig. 4d, classification performance improved with five or more generated samples per query compared to PathGen-CLIP.

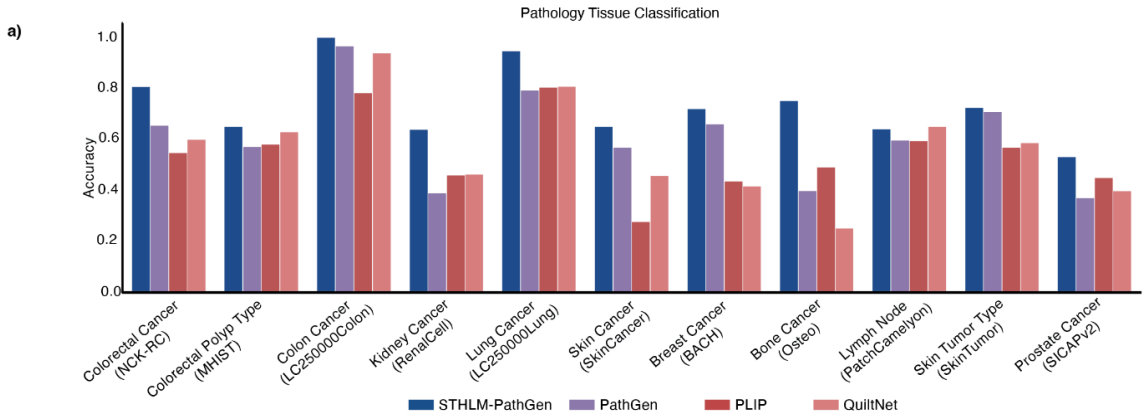

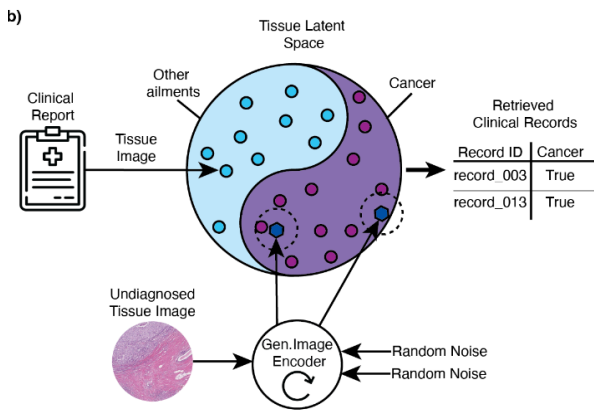
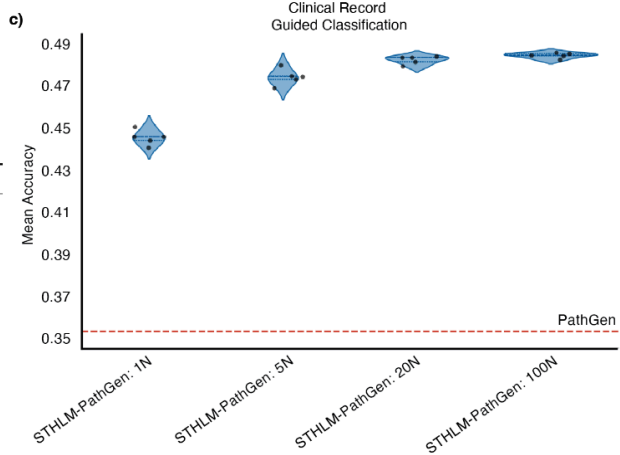

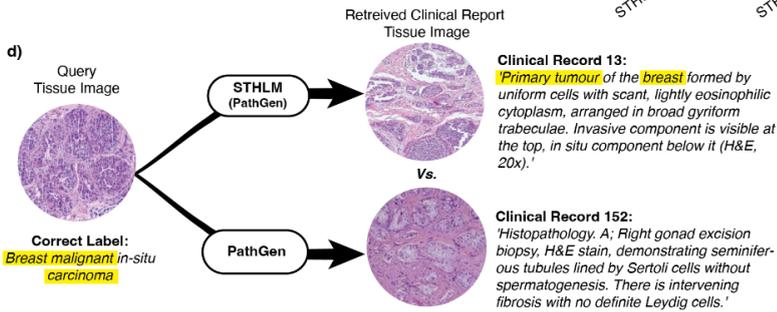
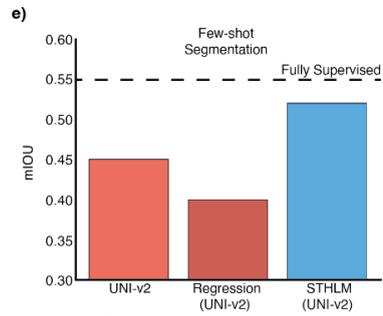

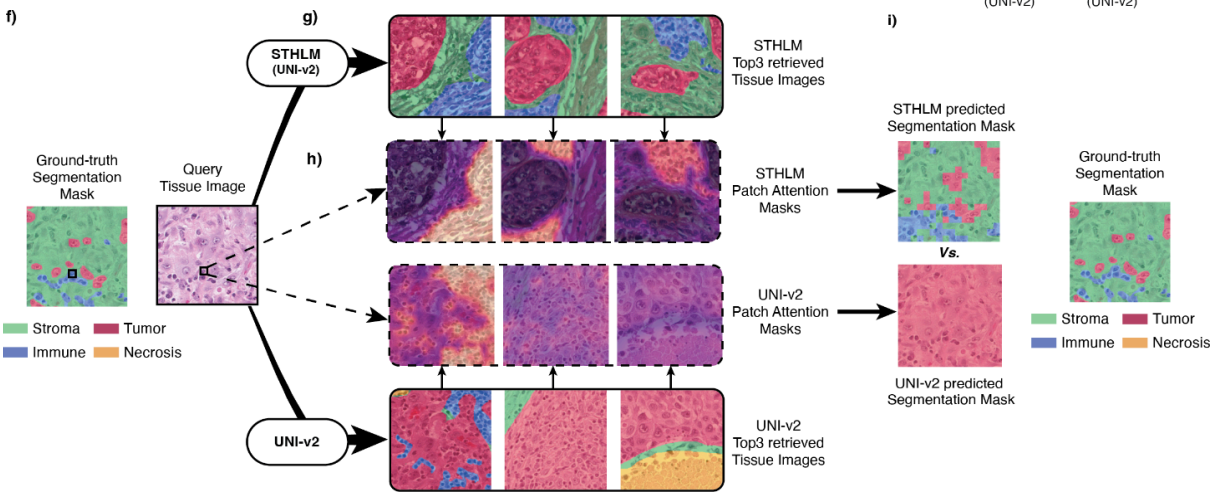

**Figure 2: Illustrations demonstrating how STHLM improves pathology foundation models for various applications**

a) Classification accuracy across 11 pathology benchmarks comparing STHLM-PathGen (200,000 samples, classifier-free guidance 0.3) to PathGen[10], PLIP[8], and QuiltNet[9]. See supplementary Fig. 5 for results across sample sizes (N) and guidance levels.

b) Histopathology images from medical records are encoded using an image encoder, with cancerous and non-cancerous samples represented as purple and blue circles, respectively. Ground truth labels (cancerous or not) are derived from clinical records. New images are classified by encoding them (blue hexagons) and performing a nearest-neighbor search against the labeled database. Retrieved labels serve as predictions, and the corresponding records provide interpretability for clinical review.

(c) Nearest-neighbor classification (3 neighbors) accuracy for histopathology images across eight tissue types (colon, breast, renal, lung, lymph node, prostate, skin, and bone), each further categorized as cancerous or non-cancerous, yielding 16 classes in total. Compared models include PathGen-CLIP and STHLM-PathGen trained with varying sample counts (N), with each N repeated five times. The figure reports mean accuracy across all classes; per-class performance is provided in Supplementary Fig. 7. Violin plots show accuracies across runs; inner marks indicate the median and quartiles (25th and 75th). Details of the experimental setup are described in the "Retrieval Strategies" section of the Methods.

(d) Representative example illustrating a case where STHLM-PathGen correctly classifies a tissue section while PathGen fails. The query image belongs to the "Breast Cancer" class, and the task is to retrieve relevant tissue images with associated clinical records to guide classification. STHLM-PathGen retrieves an image linked to a clinical report describing breast cancer, whereas PathGen retrieves an image associated with gonadal tissue, unrelated to cancer. Additional examples where STHLM-PathGen succeeds and PathGen fails are shown in Supplementary Fig. 8, and vice versa in Supplementary Fig. 9.

**e)** Comparison of UNI-v2 (red) and STHLM-UNI-v2 with 200N samples (blue) in retrieving three relevant examples for a few-shot segmentation model on the Breast Cancer Semantic Segmentation (BCSS)[22] dataset. The y-axis shows mean Intersection over Union (mIoU), an established metric for segmentation. The dashed line represents the performance of a fully supervised model trained on approximately 1,000× more images but without explainability. To isolate the impact of the generative mechanism, Regression-UNI-v2 (orange) (supplementary section "Training objective: Regression-UNI-v2") is included as a baseline, trained in the same way as STHLM-UNI-v2 (blue), highlighting the benefits of the generative approach.

**f–i)** Demonstration of the segmentation pipeline using UNI-v2 and STHLM-UNI-v2 to retrieve relevant images. See the section "Few-shot segmentation" in the supplementary to retrieve how the segmentation model was trained.

**f)** Query image with a known segmentation mask containing stroma (green), tumor (red), immune cells (blue), and necrosis (yellow).

**g)** Both UNI-v2 and STHLM-UNI-v2 encode the query image to search for relevant images to inform the segmentation. The segmentation model uses the patch embeddings (retrieved from UNI-v2 for both search methods) of these images with their corresponding class labels from the masks for attention-based classification.

**h)** A patch from the query image is classified by determining which patches it attends to in the retrieved images. This process is repeated for all patches in the query image.

**i)** Once all patches in the query image are classified, the predicted segmentation mask is reconstructed. UNI-v2 classifies the entire image as cancer, while STHLM-UNI-v2 captures more nuances, identifying immune, tumor, and stroma regions.

To assess clinical utility, we applied STHLM to the MultiCare dataset[23]. For explainable clinical-record-guided tissue classification, we retrieved nearest-neighbor images from clinical records and used their "cancer" or "non-cancer" labels and their tissue type for interpretable, verifiable predictions (Fig. 2b). In this setting, we trained an unconditional image-to-image generative model (see supplementary section) that performs local sampling based on PathGen's embeddings, resulting in STHLM-PathGen. Even with a single sample, STHLM-PathGen improves upon PathGen by mapping embeddings to more likely regions of the data manifold through local sampling (Fig. 2e). To qualitatively assess the retrieval difference, we demonstrate an example in Fig. 2d, where STHLM-PathGen successfully retrieved breast cancer tissue at a different image magnification level. In contrast, PathGen retrieved tissue related to gonads unrelated to cancer. We also built a text–image/image–text benchmark using captioned images, where STHLM variants improved retrieval across both tasks (Supplementary Fig. 4b). To reflect real-world scenarios, we also designed a multimodal retrieval task that pairs pathology images with text captions using cross-modality queries (Supplementary Fig. 4e) and compared STHLM-PathGen's retrieval performance with PathGen (Supplementary Fig. 4f).

To evaluate STHLM's potential for explainable few-shot segmentation, we benchmarked performance on the Breast Cancer Semantic Segmentation (BCSS)[22] dataset using UNI-v2[24], a state-of-the-art pathology vision encoder trained on 100 million tissue sections. We applied

STHLM to UNI-v2 to create STHLM-UNI-v2 (see supplementary section). For few-shot segmentation[25], both models retrieved the top three nearest-neighbor images with annotated masks, which were used to train a segmentation model to predict masks on query images (Fig. 2f–i). STHLM-UNI-v2 improved retrieval quality, boosting segmentation performance by nearly 15% over deterministic UNI-v2 and approaching the performance of fully supervised models trained on 1,000× more data (Fig. 2e).

In summary, we present STHLM, a generative vector search framework that enhances retrieval for tasks like RAG, classification, and interpretable diagnostics. As pathology datasets grow and the compute providers scale up beyond what has been explored in this study, both baseline and generative encoder performance are likely to improve, further strengthening STHLM. Advances in one-step generative modeling[26], non-Euclidean representations[27], and retrieval strategies may also enhance speed and accuracy, potentially reducing inference times to match those of smaller models like GTE-S while preserving the benefits of using a generative approach to vector search. A promising direction is to develop foundation models explicitly optimized for downstream generative encoding, or to jointly train both components within a unified architecture. This could further improve retrieval performance by addressing current limitations imposed by the foundation model. STHLM offers a flexible framework for advancing retrieval across biomedical applications by providing scalable test-time compute for foundation models.

**Acknowledgments**

Thanks to Lukas Käll, Reza Mirzazadeh, and other collaborators for valuable discussions. This project has received funding from the European Research Council under the European Union's Horizon 2020 research and innovation program (grant agreement no. 101021019), the Erling Persson Family Foundation, the Swedish Cancer Society, and the Swedish Research Council.

**Conflict of interest**

The authors have a patent application related to Generative Vector Search.

**Methods**

**Training objective: STHLM**

To train STHLM, we adopt Conditional Flow Matching (CFM)[28] to learn a conditional transport map from a standard multivariate Gaussian prior $N(0, I)$ to a target distribution $p(x|c)$, where $c$ is the conditioning input and $x$ is the target sample. The model learns a velocity field $\widehat{v}_\theta(x_t, t, c)$ that transports noisy samples toward the data distribution over time $t$.

For each training step, we sample:

- a noise vector $x_0 \sim N(0, I)$

- The data and query are encoded via a base encoder to produce a target sample $x_1 \sim p(x|c)$ and its corresponding conditional embedding $c$
- and a timestep $t \sim LogitNormal(0, 1)$

To reduce the distributional gap between $x_0$ and $x_1$, we apply **feature-wise standardization** to $x_1$ (zero mean, unit variance), aligning it with the scale of the Gaussian noise $x_0$.

We then construct the interpolated point:

$$x_t = tx_0 + (1 - t)x_t$$

and define the target velocity as:

$$v^* = x_1 - x_0$$

The CFM loss minimizes the mean squared error between the predicted and target velocities:

$$L_{CFM}(\theta) = E_{x_0, x_1, t}\left[||\widehat{v}_\theta(x_t, t, c) - v^*||^2\right]$$

When $\widehat{v} = v^*$, it can be shown that the marginal distribution of $\widehat{x}_1$ is obtained by integrating the learned velocity field from $x_0 \sim N(0, I)$ corresponds to the target distribution $p(x|c)$[28].

Where:
- θ is the learnable parameters of the model
- $\widehat{v}_\theta(x_t, t, c)$ is the velocity predicted by the network with parameters θ, given the state $x_t$ at time $t$ and condition $c$.

To improve diversity and sample quality, we deployed classifier-free guidance. During training, 10% of the condition *c* were dropped out, allowing the model to learn to handle both unconditional and conditional generative modeling.

For the image-to-image STHLM-PathGen trained an unconditional model, omitting the condition *c*.

**Conditional inference: STHLM**

STHLM generates target embedding conditioned on the query at inference time by integrating the learned velocity field $\hat{v}_\theta(x_t, t, c)$ in reverse, from a standard multivariate Gaussian to the target distribution, using an Euler ODE solver. Starting from a sample $x_0 \sim N(0, I)$, the solver iteratively updates the sample by following the velocity field $\hat{v}_\theta(x_t, t, c)$, conditioned on the input *c*. For all benchmarks, we used 10 denoising steps with the Euler ODE solver, unless stated otherwise. We also evaluated the effect of the number of Euler steps on retrieval performance, as shown in Supplementary Fig. 6.

At inference time, when doing classifier-free guidance, we interpolate between the conditional and unconditional velocity estimates:

$$\hat{v}_\theta(x_t, t, c) = \hat{v}_{uncond, \theta}(x_t, t, 0) + \lambda(\hat{v}_{cond, \theta}(x_t, t, c) - \hat{v}_{uncond, \theta}(x_t, t, 0))$$

where $\hat{v}_{uncond, \theta}(x_t, t, 0)$ is the unconditional velocity field, $\hat{v}_{cond, \theta}(x_t, t, c)$ is the conditional velocity field given the condition *c*, and $\lambda$ is the guidance scale. Lower guidance scale $\lambda$ value places less emphasis on $c$ when generating the target embedding, allowing for greater diversity. We used classifier-free guidance in all benchmarks with a scaling factor of 1.0, except for the benchmark shown in Fig. 2a, where we used a scaling factor of 0.3. This benchmark differs from the others in that it involved classifying up to 10,000 samples, where increased diversity appeared to improve performance. In Supplementary Fig. 5, we investigate how varying the scaling of classifier-free guidance affects retrieval performance.

**Unconditional inference: STHLM**

In cases where no conditioning information was available, such as in the clinical report–guided classification setup for STHLM-PathGen, we employed local sampling. In this approach, an existing embedding is perturbed with noise at a given time point $t$, and then denoised to recover likely samples near the original data point. In other words, local sampling draws samples from the neighborhood of a noisy version of the image embedding, allowing exploration around a specific embedding rather than complete unconditional generation.

We empirically determined suitable starting time points by testing $t$ values between 0.5 and 0.9. For STHLM-PathGen, a value of $t = 0.6$ provided a good balance between sample diversity and specificity, whereas STHLM-UNI-v2 (still using conditional modelling, but using local sampling) performed best with $t = 0.8$, where the higher noise level was compensated by the presence of a condition embedding.

**Training objective: Regression-UNI-v2**

For Regression-UNI-v2, the goal is to predict a target embedding $x$ conditioned on $c$ by learning a function $f$ such that $x = f_\theta(c)$. To maintain architectural consistency with STHLM—which uses a conditional network that takes both a noise vector and a condition—we modify the input to be $x = f_\theta(0, c)$, where the zero vector replaces the stochastic component. This allows us to reuse the same model architecture while treating the regression task deterministically.

We adopt a loss structure similar to that used in STHLM, using a mean squared error (MSE) loss. The MSE loss is defined as:

$$L_{MSE}(\theta) = E_x \left[ ||f_\theta(0, c) - x||^2 \right]$$

**Training Datasets**

To train STHLM–NV-Embed-v2, we utilized the training portion of NFCorpus[18], which comprises approximately 135,800 query–document relevance pairs. We further incorporated 10,800 synthetic query–document pairs from the *BeIR/nfcorpus-generated-queries* dataset available on Hugging Face. Furthermore, we used 1,409 expert-curated corpus texts from the SciFact training dataset[29]. We also generated 10 synthetic queries for each corpus text with *ChatGPT-4o*-mini. We did the same query generation strategy for BioASQ[19], which contains approximately ~40,200 corpus texts. To augment more queries per corpus, we also did random linear interpolation between two random queries per corpus text in the embedding space.

For STHLM-PathGen, we used 207,000 paired samples from the PathCap dataset[30] and an additional 100,000 samples from QuiltNet[9], similar to PathGen[10], filtered to include only those with a "dirtiness" score below 0.1, determined by heuristics as described in reference[31]. Access to the PLIP dataset[8] was unavailable. While PathGen[10] provides a large dataset, its authors note that only a subset is pathologically relevant; since this subset could not be reliably identified, the dataset was excluded from training. Given the limited data volume (307,000 samples), we generated five synthetic queries per text sample using ChatGPT-4o-mini to increase training diversity. During training, batches were sampled using a 1:1 ratio of real to synthetic examples. Additionally, in the multimodal setting, we applied a data augmentation strategy in which the roles of text and image were swapped to expand the training data further. This approach leveraged that the base encoder, PathGen, was trained on the same dataset, ensuring reasonably well-aligned text–image pairs. As a result, using the text or image as both the conditioning input and the target (for example, creating (text, text) or (image, image) pairs) provided additional, meaningful supervision during model training.

For the image-to-image model STHLM-PathGen used for clinical report guided classification, we used a collection of publicly available histopathology datasets: NCK-RC[32], MHIST[33], LC25000 (colon and lung)[34], BACH[35], SICAPv2[36], Osteo[37], PatchCamelyon[38], SkinCancer, SkinTumor[39], and RenalCell[40].

| Dataset | Samples | Classes | Class distribution |
|---|---|---|---|
| NCK-RC | 6 333 | 8 | 0.21/0.05/0.1/0.16/0.09/0.11/0.07/0.19 |
| MHIST | 977 | 2 | 0.63/0.37 |
| LC25000Colon | 10 000 | 2 | 0.5/0.5 |
| RenalCell | 36 687 | 5 | 0.027/0.36/0.25/0.23/0.145 |
| LC25000Lung | 15000 | 3 | 0.33/0.33/0.33 |
| SkinCancer | 28 039 | 12 | 0.032/0.023/0.029/0.015/0.002/0.071/0.011/0.093/0.016/0.12/0.17/0.091/0.123 |
| BACH | 400 | 4 | 0.25/0.25/0.25/0.25 |
| SICAPv2 | 2122 | 4 | 0.30/0.19/0.40/0.10 |
| Osteo | 1091 | 3 | 0.49/0.24/0.27 |
| PatchCamelyon | 32768 | 2 | 0.50/0.50 |
| SkinTumor | 8 851 | 4 | 0.39/0.30/0.10/0.20 |

For STHLM-UNI-v2, we used ~3,000 image-mask pairs from the BCSS[22] dataset. We made a binary vector for each mask, 1 if a class exists in the mask, and 0 otherwise. This binary was then used as a condition for the generative model to retrieve its corresponding image.

**Evaluation Datasets**

To benchmark the STHLM-NV-Embed-v2 shown in *Fig. 1g*, we utilized the SciFact dataset[29], which consists of 1,409 corpus texts. The validation set contains 450 claims, each linked to one or more of the corpus texts. The NFCorpus[18] validation dataset includes approximately 12,334 query-medical judgment pairs, while BioASQ contains roughly 4,700 question-answer pairs.

To evaluate STHLM in comparison to pathology foundation models, as shown in *Fig. 2f*, we used a suite of publicly available datasets, including NCK-RC[32], MHIST[33], LC25000 (colon and lung)[34], BACH[35], SICAPv2[36], Osteo[37], PatchCamelyon[38], SkinCancer, SkinTumor[39], and RenalCell[40].

To assess STHLM's ability to improve multi-label classification, as shown in Supplementary Fig. 1d, we used a BCSS[22] containing four primary classes: tumor, necrosis, stroma, and inflammation. Images were assigned a class label if more than 1% of their area belonged to that class. This resulted in a dataset of 6,000 images, each potentially associated with multiple labels. To determine the similarity threshold for classification, we used 20% of the data as a tuning set (to optimize the F1-score) and evaluated performance on the remaining 80%.

To evaluate STHLM's ability to infer the presence of cancer and tissue type in histopathology images based on textual evidence, as shown in Fig. 2i, we extracted 3,002 image–captions pairs from clinical reports in the MultiCare dataset[23]. Because the MultiCare dataset contained relatively few breast and prostate tissue samples, we supplemented it with 793 breast and 335 prostate image–caption pairs from the PathCap dataset[30]. We used ChatGPT-4o-mini to classify the tissue type associated with each image–caption pair and to determine whether it was related to cancer. Across tissue types (after removing the ones ChatGPT-4o-mini couldn't determine), the dataset comprised bone (295 cancer, 230 non-cancer), breast (169 cancer, 658 non-cancer), colon/colorectal (81 cancer, 70 non-cancer), kidney/renal (134 cancer, 80 non-cancer), lung (117 cancer, 92 non-cancer), lymph node (97 cancer, 115 non-cancer), prostate (120 cancer, 221 non-cancer), and skin (325 cancer, 122 non-cancer). For balanced evaluation, we used a dataset comprising 200 samples per class (from NCK-RC[32], MHIST[33], LC25000 (colon and lung)[34], BACH[35], SICAPv2[36], Osteo[37], PatchCamelyon[38], SkinCancer, SkinTumor[39], and RenalCell[40]) covering both cancer and non-cancer ("other") cases across eight tissue types: bone, breast, colon/colorectal, kidney/renal, lung, lymph node, prostate, and skin (16 classes in total).

To evaluate retrieval performance across modalities, as shown in Supplementary Fig. 3b, we used the MultiCare dataset[23] for text-to-image and image-to-text retrieval tasks. For text-to-text retrieval, we constructed a corpus of 72,581 clinical record descriptions. From this corpus, we randomly selected 100 and 500 descriptions and generated five synthetic queries per description using ChatGPT-4o-mini. This resulted in query–description pairs that formed the validation and test sets of the clinical text retrieval benchmark.

For image-to-text and text-to-image retrieval, we utilized 2,821 unique captions associated with 3,012 histology images. These same data were also used in the multi-query benchmark (Supplementary Fig. 3f), where both the caption and the image served as query inputs for retrieving the correct caption–image pair. For each pair, either the image or the caption was used as the query, and retrieval performance was evaluated by checking whether the corresponding counterpart was successfully identified.

For STHLM-UNI-v2, we had approximately 1,000 image–mask pairs from the BCSS[22] dataset as evaluation data. However, since the masks should be "unknown" we trained a simple 2D conv-network multilabel classifier on the training tissue sections. We used this classifier to predict the classes on the evaluation tissue sections which were then used as conditions for Fig. 2e.

**Retrieval Strategies**

The process of retrieving and ranking results after generating samples with the conditional generative encoder is a critical component of generative vector search. In this work, we employed k-nearest neighbors retrieval with 3 neighbours (if not stated otherwise). While these naive approaches provide a strong baseline, exploring more advanced retrieval and ranking strategies would be interesting for future work.

For the pathology classification benchmarks shown in Fig. 2a, the task differs from the others. Here, we classify over ten thousand examples, whereas the others focus on ranking top results or classifying a few hundred samples. In this setting, getting enough samples per class for k-nearest neighbors may be computationally infeasible. Instead, we trained a simple conditional classifier on our generated samples. Specifically, we applied CORrelation ALignment (CORAL) to align the generated data with the target (test) samples[41]. A class-conditioned von Mises–Fisher classifier[42] is then trained on the aligned synthetic data and used to classify the test samples. To obtain results from the other models, we utilized the evaluation scripts provided by Quilt1M[9] (available at [https://github.com/wisdomikezogwo/quilt1m](https://github.com/wisdomikezogwo/quilt1m)) and generated query embeddings from PathGen using the same pipeline we employ for STHLM.

We used one nearest neighbor per class embedding for the multi-label classification task demonstrated in Supplementary Fig. 3d. For each image, we sampled the image embeddings, calculated its distance to each class embedding, and found the closest class embedding. We then summed up the number of times each class embedding was closest and divided it by the number of samples to form a class distribution per image, which was then used to assign multi-label predictions.

**CORrelation ALignment (CORAL)**

CORAL can align synthetic data with real target data, thereby mitigating domain shift[43]. Given real data $X_r \in R^{NxD}$ and synthetic data $X_s \in R^{SxD}$, we first normalize both datasets so that all samples lie on the unit sphere, i.e., $\left\Vert x_i \right\Vert_2 = 1$. To perform CORAL, both datasets are first centered by subtracting their respective means:

$$\overline{X}_s = X_s - \mu_s, \overline{X}_r = X_r - \mu_r$$

Next, the covariance matrices are computed:

$$C_s = cov(\overline{X}_s), C_r = cov(X_r):$$

The synthetic data is then linearly transformed to match the second-order statistics of the real data:

$$X_s^{cov-aligned} = \overline{X}_s C_s^{-\frac{1}{2}} C_r^{\frac{1}{2}}$$

Then, we re-center the transformed data by adding the mean of the real data to align the first-order statistics:

$$X_s^{aligned} = X_s^{cov-aligned} + \mu_r$$

This transformation aligns the covariance structure of the synthetic data to that of the real domain. The aligned features $X_s^{aligned}$ can then be used in downstream tasks alongside $X_r$, with reduced distributional discrepancy.

When using $X_s^{aligned}$ for von Mises–Fisher (vMF) classification, we must ensure that the features lie on the unit hypersphere. This is achieved by normalizing each sample:

$$X_s^{aligned} = \frac{X_s^{aligned}}{\|X_s^{aligned}\|_2}$$

**Conditional classification with von Mises-Fisher Distribution**

The feature vectors used in this paper are all derived from representation learning models in which the embeddings are constrained to lie on the unit hypersphere. This makes the von Mises–Fisher (vMF) distribution a more natural choice than, for example, a Gaussian distribution for modeling class-conditional densities[42].

For each class *c*, we model its distribution using a mean direction vector $\mu_c \in S^{D-1}$ and a concentration parameter $\kappa_c > 0$. The likelihood of an embedding $x \in S^{D-1}$ under a class *c* is given by:

$$P(x|c) = C_D(\kappa_c) e^{\kappa_c \mu_c^T x}$$

where the normalizing constant is:

$$C_D(\kappa) = \frac{\kappa^{\frac{D}{2}-1}}{(2\pi)^{\frac{D}{2}} I_{\frac{D}{2}-1}(\kappa)}$$

and $I_v(\cdot)$ is the modified Bessel function of the first kind of order $v$. To compute $log(C_D(\kappa))$ stably, we use a scaled form:

$$log(I_v(\kappa)) \approx log(e^{-\kappa} \cdot I_v(\kappa)) + \kappa,$$

which avoids overflow at large $\kappa$.

Class parameters $\mu_c$ and $\kappa_c$ are estimated from the generated synthetic data for each class $X_c$. We compute the mean direction as:

$$\mu_c = \frac{1}{n} \sum_{i=1}^{n} x_i$$

We need to ensure the mean is normalized:

$$\bar{\mu}_c = \frac{\mu_c}{\|\mu_c\|_2}$$

To estimate the concentration parameter, we use:

$$\kappa_c \approx \frac{\bar{R}_c (D - \bar{R}_c^2)}{1 - \bar{R}_c^2}$$

where $\overline{R}_c$ is the mean resultant length of $X_c$ defined as:

$$\overline{R}_c = \left\| \frac{1}{n} \sum_{i=1}^{n} x_i \right\|$$

To classify on real data $x \in X_R$, we compute the log-posterior over classes using Bayes' rule:

$$log(p(c|x)) \sim \kappa_c \overline{\mu}_c^T x + log(C_D(\kappa_c)) + log(p(c))$$

We assume a uniform prior $p(c)$.

**Performance Metrics**

Across benchmarks, we used NDCG@10, accuracy, F1-score, and mIOU (mean Intersection over Union) to evaluate model performance.

Let $y$ be the true labels and $\hat{y}$ be the predicted labels:

- Accuracy is the proportion of correct predictions:

$$Accuracy = \frac{1}{N} \sum_{i=1}^{N} 1(y_i = \hat{y}_i)$$

- F1-score is the harmonic mean of precision and recall:

$$F1 = 2 \frac{Precision * Recall}{Precision + Recall}$$

- mIOU was used to compute segmentation performance. For each class $c$, the Intersection over Union (IoU) and mIOU is calculated as:

$$IOU = \frac{|Prediction_c \cap Ground\ Truth_c|}{|Prediction_c \cup Ground\ Truth_c|}$$

$$mIOU = \frac{1}{C} \sum_{c=1}^{C} IOU_c$$

- NDCG@10 (Normalized Discounted Cumulative Gain at rank 10) measures the quality of a ranked list by considering both the relevance of retrieved items and their positions in the ranking. It is defined as:

$$NDCG@10 = \frac{DCG@10}{IDCG@10}$$

where

$$DCG@10 = \sum_{i=1}^{10} \frac{2^{rel_i} - 1}{\log_2(i+1)}$$

and $rel_i$ is the relevance score of the item at rank $i$, typically binary (0 or 1) for relevance-based benchmarks. IDCG@10 is the ideal DCG—i.e., the maximum possible DCG@10 for a perfect ranking.

NDCG@10 ranges from 0 to 1, with higher values indicating better ranking quality. It is particularly useful in retrieval tasks where the position of relevant documents is essential.

**Model Architecture**

All STHLM variants use hypernetworks[44] combined with low-rank adaptation[45]. The hypernetworks use a HyperLinear component that is composed of a weight matrix and dynamic low-rank corrections. The core building block is a HyperLinear layer, which augments a standard linear transformation with dynamic, low-rank corrections conditioned on the input embedding $c \in R^{d_{cond}}$. The HyperLinear operation is defined as:

$$y = s(c) \odot (Wx) + \alpha[V(c)U(c)^T]x + b(c)$$

Where:

- $\odot$ denotes the element-wise product

- α is a learnable parameter initialized to a small value for training stability

- $x \in R^{d_{in}}$ is the input,

- $W \in R^{d_{out} \times d_{in}}$ is a learned weight matrix,

- $U(c) \in R^{d_{in} \times r}$, $V(c) \in R^{d_{out} \times r}$ are low-rank matrices generated from *c* via a small multi-layer perceptron (MLP),

- $b(c) \in R^{d_{out}}$ is a learned, condition-dependent bias generated from *c* via a small MLP,

- and $s(c) \in R^{d_{out}}$ is a learnable scaling modifier generated from *c* via a small MLP.

The rank $r$ is chosen such that $r < min(d_{out}, d_{in})$ and ensures the correction remains low rank and efficient. To stabilize training during early optimization, we initialize all dynamic components—including $U(c), V(c), b(c), and\ s(c)$—to zero or the identity matrix. This ensures that the HyperLinear layer behaves like a standard linear projection at initialization.

We define a HyperMLP as a two-layer MLP composed of consecutive HyperLinear layers, with a GELU activation in between. This module is used within a residual block of the following form:

$$F(x, c) = x + HyperMLP(LayerNorm(x), c)$$

The final model is constructed by stacking multiple such residual blocks, allowing the model to adapt flexibly to the conditioning input $c$ while benefiting from the stability and efficiency of low-rank modulation.

**Model Hyperparameters**

In all STHLM model variants, HyperMLPs serve as the core architectural building blocks. The input is first projected into a higher-dimensional latent space. The conditioning embedding is batch-normalized and linearly transformed to align it with the model's internal representation.

To incorporate temporal information, we use a sinusoidal positional encoding with a base frequency $\frac{1}{10000}$ to vectorize the timestep, which is passed through a dedicated multilayer perceptron (MLP) to produce a learned time embedding. This time embedding is concatenated with the processed condition embedding, and the combined representation is projected into a unified conditional embedding.

The final conditional embedding modulates each HyperMLP block throughout the model stack. Each HyperMLP block is condition-aware, enabling the model to dynamically adapt its behavior based on context and time. The output from the final HyperMLP layer is then projected to match the desired output dimensionality.

We reduced NV-Embed-v2's embeddings from 4096 to 384 dimensions via PCA, matching GTE-S dimensionality for fair comparison. Similarly, UNI-v2 embeddings were reduced from 1536 to 64 dimensions. Higher-dimensional projections yielded negligible performance improvements while increasing computational costs, suggesting these dimensions capture the essential semantic structure for the tested benchmarks.

| Modality | Input (dim.) | Time (dim.) | Hidden (dim.) | Condition (dim.) | Layers | Lora rank | Model parameters |
|---|---|---|---|---|---|---|---|
| STHLM-NV-Embed (v2) | 384 | 768 | 768 | 384 | 4 | 32 | 496M |
| STHLM-PathGen (text-to-image) | 512 | 512 | 768 | 768 | 3 | 32 | 372M |
| STHLM-PathGen (image-to-text) | 512 | 512 | 768 | 768 | 3 | 32 | 372M |
| STHLM-PathGen (image-to-image) | 512 | 512 | 768 | 768 | 3 | 32 | 332M |
| STHLM-UNI-v2 | 64 | 32 | 256 | 6 | 1 | 16 | 1.2M |

**Training Hyperparameters**

When training

| Modality | Optimizer | Lr | Wd | Cosine Warmup (steps) | Batch Size | Epochs | Total steps |
|---|---|---|---|---|---|---|---|
| STHLM-NV-Embed (v2) | AdamW | 1e-4 | 1e-5 | 500 | 1024 | 100 | 98.6k |
| STHLM-PathGen (text-to-image) | AdamW | 1e-4 | 1e-5 | 500 | 512 | 100 | 124.8k |
| STHLM-PathGen (image-to-text) | AdamW | 1e-4 | 1e-5 | 500 | 512 | 100 | 124.8k |
| STHLM-PathGen (image-to-image) | AdamW | 1e-4 | 1e-5 | 500 | 1024 | 200 | 25.6k |
| STHLM-UNI-v2 | AdamW | 1e-4 | 1e-5 | 500 | 256 | 20 | 69.5k |

**Code availability**

The code for STHLM can be found at https://github.com/ekvall93/STHLM

**Data availability**

- The SciFact dataset[29] can be accessed https://huggingface.co/datasets/allenai/scifact.
- BioASQ[19] dataset: https://huggingface.co/datasets/rag-datasets/rag-mini-bioasq
- NFCorpus[18] dataset: https://huggingface.co/datasets/BeIR/nfcorpus

- Evaluation datasets used in pathology benchmarks were obtained using scripts from the Quilt-1M codebase[9], available at https://github.com/wisdomikezogwo/quilt1m.
- Training datasets for pathology benchmarks include PathCap[30](https://huggingface.co/datasets/jamessyx/PathCap) and Quilt-1M[9](https://github.com/wisdomikezogwo/quilt1m).
- Clinical report data were sourced from the MultiCaRe dataset[23], available at https://github.com/mauro-nievoff/MultiCaRe_Dataset/tree/main.
- The BCSS multi-label segmentation dataset[22] was downloaded from Kaggle https://www.kaggle.com/datasets/whats2000/breast-cancer-semantic-segmentation-bcss.

**Prompts for synthetic data generation**

The generate synthetic queries for STHLM-NVEMBED-v2 we used the following prompt:

*"""Your task is to generate a diverse list of 10 queries for a given text.*

*The queries should be delivered as items within a single JSON object with one key, "queries", which holds a list of 10 unique queries. Only return the JSON object—do not include any other explanation or content.*

*# Output Format*

*Return a single JSON object containing one key: "queries". The value should be an array of exactly 10 queries, each as a string. No additional fields, wrapping, or explanation.*

# Examples

```
{
  "queries": [
    "Is the protein Papilin secreted",
    "Which miRNAs could be used as potential biomarkers for epithelial ovarian cancer?",
    "Which proteins participate in the formation of the ryanodine receptor quaternary macromolecular complex?",
    "What is the role of the protein Papilin in the cell cycle?"
  ]
}
```

Use Examples to get a variety of questions that is returned JSON object
"""

To augment the pathology training data with synthetic examples, we employed ChatGPT-o4-Mini to generate five novel queries for each caption in the PathCap and Quilt-1M datasets. The prompting template was as follows:

"""

You are a medical data assistant. I will give you a sentence or paragraph describing histopathological or biomedical findings.

*Your task is to extract the key concepts (e.g., tissue structures, injuries, conditions, etc.) and generate **5 descriptive queries** using natural language templates suitable for image retrieval. Use relevant medical synonyms where helpful.*

*Format your output as a Python list of strings.*

*Templates you may use (but are not limited to):*

* *"a histopathology slide showing {c}"*

* *"histopathology image of {c}"*

* *"pathology tissue showing {c}"*

* *"microscopic image highlighting {c}"*

* *"presence of {c} tissue on image"*

* *"tissue section showing {c}"*

* *"cardiac tissue sample with {c}"*

*Avoid redundancy across queries, and aim to cover different combinations or synonyms of the findings.*
"""

To construct evaluation and training datasets from the clinical reports in the MultiCare dataset, we first cleaned the clinical summaries using the following prompt:

"""

*You are a Python API that receives clinical text that describes an image. Clean up the text: remove weird characters, parentheses with technical details, and return a plain cleaned-up string with no extra quotes.*

*Here is the clinical text:*

*{text_mass}*

*"""*

Subsequently, we generated search queries for the cleaned texts using the following prompt:

*"""*

*You are a Python API that receives clinical text. I want you to generate 5 diverse clinical queries that would help the clinician find similar text mass in a database. Only output a Python list of clinical queries, so it's easy to parse.*

*Here is the clinical text:*

*{text_mass}*

*"""*

To classify the tissue type and determine whether it was related to cancer, we used the following prompt:

*"""*

"You are a medical assistant. For each medical caption, do two things:\n"
    "1) Identify the primary anatomical site involved, using ONLY these IDs:\n"

"   0: bone; 1: breast; 2: colon/colorectal; 3: kidney/renal; "

"   4: lung; 5: lymph node; 6: prostate; 7: skin.\n"

"   If the site is unclear or mentions multiple unrelated sites with no clear focus, use -1.\n"

"2) Determine if it is related to cancer (malignancy), where cancer=1 if the caption clearly involves "

"   malignancy (e.g., carcinoma, sarcoma, melanoma, lymphoma, leukemia, metastasis, 'invasive', or 'in-situ carcinoma'); "

"   otherwise cancer=0.\n"

"Output ONLY a JSON object like: {\"site_id\": <int>, \"cancer\": <0 or 1>} with no extra text."

**Few-shot segmentation**

For the few-shot segmentation task, we used UNI-v2 as the base model, which is trained using a Self-Distillation with No Labels (DINO)[46]-based architecture, to develop STHLM-UNI-v2. To train STHLM-UNI-v2, we trained a conditional generative model to denoise UNI-v2 embeddings conditioned on multilabel encodings of the segmentation masks (where each class was encoded as 1 if present in the mask and 0 otherwise). During inference on unseen images, a simple regressor was used to predict the multilabel encoding. The predicted encoding was then combined with local sampling of the image embeddings (at $t = 0.8$) to generate samples and retrieve the most likely corresponding images from the database.

For both UNI-v2 and STHLM-UNI-v2, ten images were retrieved per query; seven were used for training and three for evaluation of the segmentation model. For each retrieved image, UNI-v2 (for both UNI-v2 and STHLM-UNI-v2) is used to retrieve *16x16* patch embeddings, and the majority class for the patch is assigned to the corresponding patch embedding. All training patch-embedding is used as prototypes. During training, image patch embeddings are linearly transformed with trainable weights, likewise all prototypes are linearly transformed with trainable weights. The cosine distance is then computed between all patch embeddings and all prototypes (prototypes originating from the same image used for training is masked), and a softmax with a learnable temperature is used to compute the attention. The attention of the prototypes can then be used to compute the logits of the most probable class. The image-wise bias term is used for all prototypes originating from an image, to remove attention from certain images. The model stops the training when validation images stop improving after 2 epochs.

To evaluate the query image, the bias term is used to select the top 3 image prototypes, which are then used to classify the query image patch embeddings.

**Vector Database Capacity Limitations and *Generative Modelling***

To illustrate the fundamental challenge of vector database capacity, we constructed a simplified example using histology images. Despite straightforward data and simple classification tasks, vector databases face a capacity limit—similar to a "pigeonhole principle" where too many items must be squeezed into limited space. When this happens, embeddings from different classes begin to overlap, making retrieval difficult (supplementary Fig. 1a). Traditional approaches that represent each query or class with a single embedding struggle under these constraints (supplementary Fig. 1b). In contrast, a generative approach that samples multiple embeddings per query maintains better separation and retrieval accuracy (supplementary Fig. 1c), especially when representational capacity is limited. This advantage becomes even more pronounced in real-world biomedical applications where queries naturally span multiple concepts and the embedding space must accommodate vast, complex data. Generative modeling offers a natural solution to these challenges.

To get a dataset for this experiment, we combined images from NCK-RC[32] (7 classes), RenalCell[40] (5 classes), and SkinCancer[39] (16 classes) datasets. For each dataset, 300 data points were used for training and 100 for testing. Hierarchical categorizations included: tissue type (colon: 2100/700 train/test, kidney: 1500/500, skin: 4200/1400), compartment (epithelium: 1800/600, stroma: 900/300, immune: 300/100, vascular: 600/200, acellular: 900/300, neural: 300/100, other: 3000/1000), and cancer status (cancer: 1500/500, non-cancer: 6300/2100).

All images were encoded using PathGen to generate vector representations, then reduced to varying dimensionalities using principal component analysis (PCA). An MLP classifier was optimized to map image encodings to learnable class prototypes (28 + 3 + 7 + 2), with class weighting until convergence on test data. Final classification used the cosine distance between image encodings and class prototypes. For the generative approach, a CFM conditional generative model, which used a simple MLP with 256 hidden dimensions, was trained on the training data to generate new samples. These generated samples were then used to train a classifier for test data classification.

**Hardware**

All experiments were conducted on an NVIDIA A100-SXM4-80GB GPU.

**Supplementary Figures**

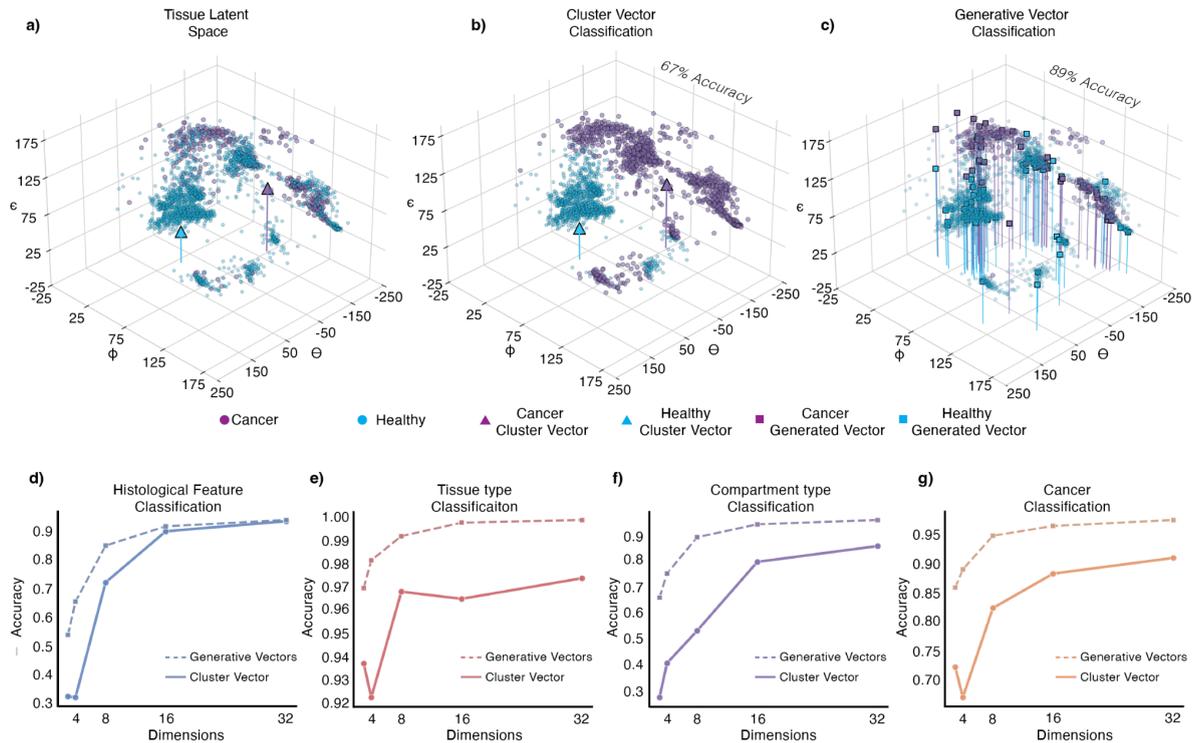

**Supplementary Figure 1: More comparison of classification accuracy between cluster embedding-based and generative modelling across dimensionalities and tasks.**

**a–c)** Demonstrates how generative embeddings improve retrieval when database capacity is constrained.

**a)** A simplified vector database with histology images from multiple tissues and conditions. Purple points = cancer samples; blue points = healthy samples. Even in this straightforward setup, samples from different classes overlap when forced into a low-dimensional space (here visualized in 3D). Triangles show the single representative embedding for each class.

**b)** Classification accuracy using single embeddings per class (triangles). Overlapping samples lead to classification errors.

**c)** Instead of single points, a generative model creates multiple embeddings per class (squares), better capturing the spread of samples within each category.

**d-g)** Generative modelling outperforms cluster embedding-based classification across all category hierarchies. The performance gap between generative modelling and cluster embedding is especially pronounced at low dimensions, with cluster embedding gradually improving as the number of embeddings increases, as expected. The cancer classification (**g**) corresponds to the labels used in (**a–c**), while each data point also contains labels from the other hierarchical classification levels (**d–f**).

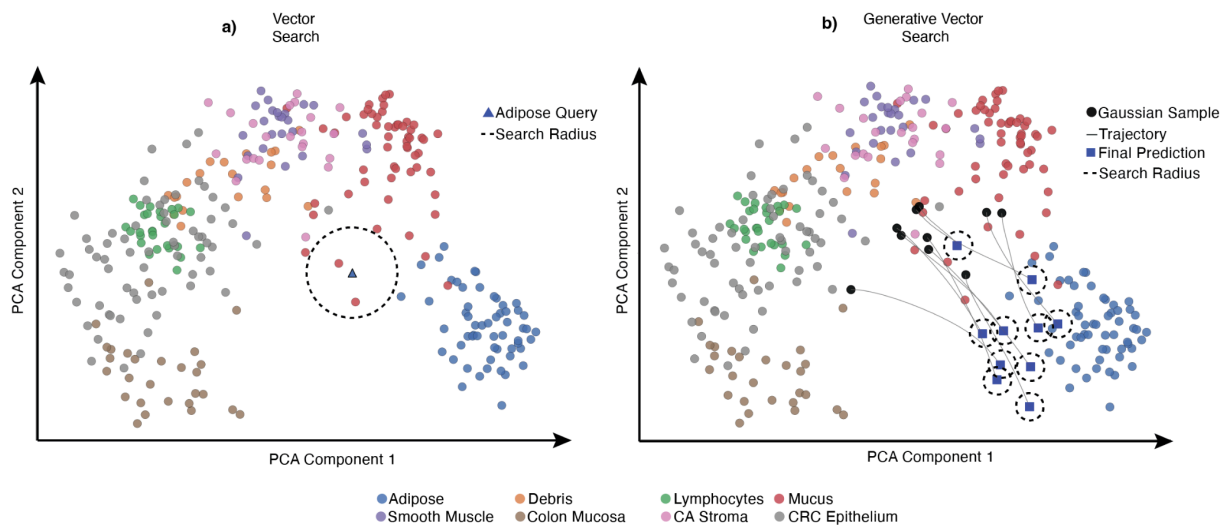

**Supplementary Figure 2:** PCA of latent space from the NCK-RC dataset[32] (100,000 H&E-stained images, eight tissue classes). Embeddings are shown for a discriminative model (a) and a conditional generative model (b), allowing for a direct comparison of class latent matching in PCA space.

a) The triangle represents a single point estimate from the query "Image showing Adipose tissue" encoded by a text encoder. While distances in the 2D PCA space do not precisely reflect distances in the original embedding space, the plot illustrates how a single embedding may lie closer to irrelevant samples than to relevant ones.

b) STHLM uses Conditional Flow Matching (CFM)[28] to transform random samples (black circles) along trajectories (black lines) into query-conditioned embeddings (blue squares), improving retrieval relevance.

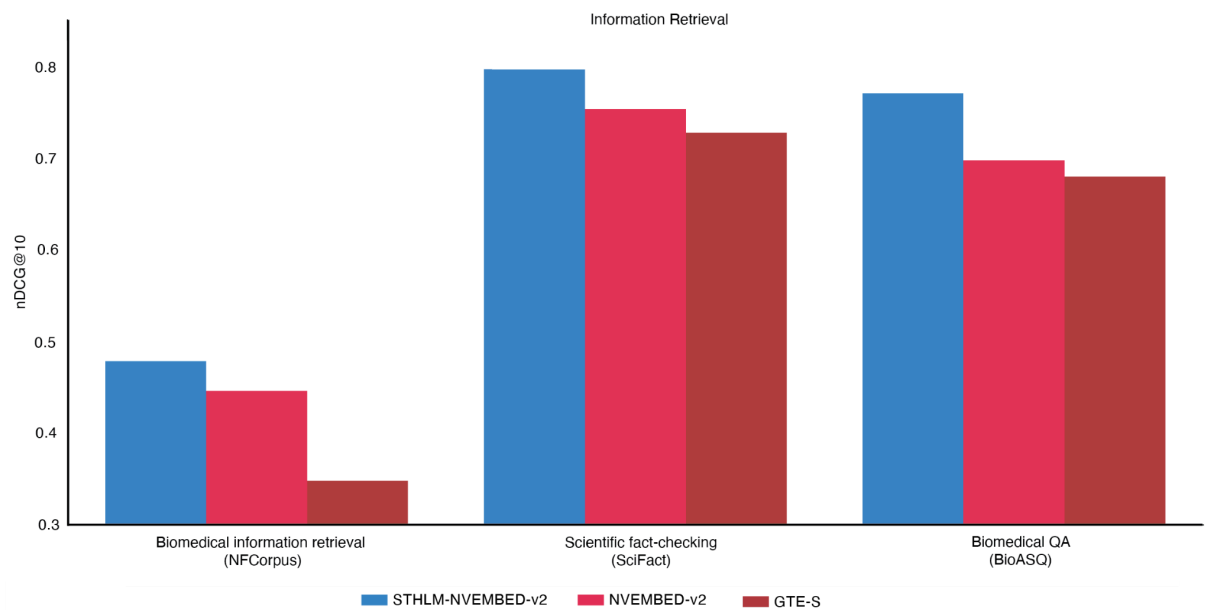

**Supplementary Figure 3:** Biomedical information retrieval comparison between STHLM-NVEMBED-v2 (blue) using 60N samples, NV EMBED-v2, and GTE-S, on three different biomedically relevant dataset: NFCorpus, SciFact, and BioASQ.

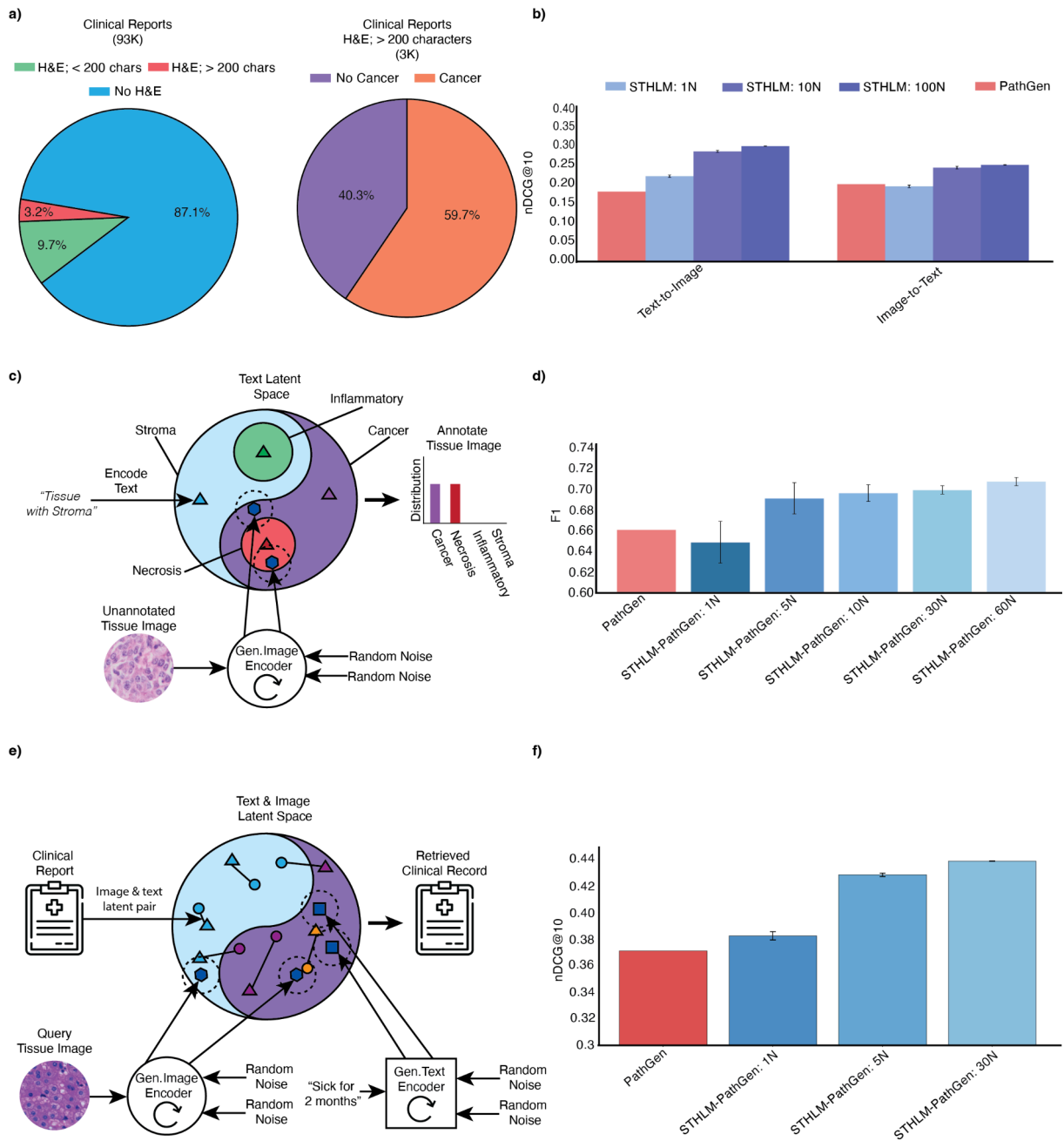

**Supplementary Figure 4: MultiCare dataset overview and performance of multimodal retrieval benchmarks** (All bar plots display mean performance with error bars denoting ±1 standard deviation).

**a)** Data composition of the MultiCare dataset. Left, pie chart: The dataset comprises 93,000 clinical reports, all used for text-to-text retrieval. Among these, ~12,100 reports include Hematoxylin and Eosin (H&E) stained images. Of the H&E-associated reports, ~9,000 have captions shorter than 200 characters (H&E < 200 chars), and ~3,000 have captions exceeding 200 characters (H&E > 200 chars). Only the latter subset was used for image-to-text and image-to-image tasks. Right, pie chart: Within the H&E > 200 chars subset, ~1,800 reports were cancer-related (Cancer), while ~1,200 described other conditions (No Cancer).

**b)** Multimodal retrieval benchmarks (MultiCare dataset): comparing PathGen-CLIP vs. STHLM-PathGen for text-to-image/image-to-text, across sample counts (N).

**c)** For multi-label classification, a text encoder embeds each class label—"cancer" (purple triangle), "necrosis" (red triangle), "inflammatory" (green triangle), and "stroma" (blue triangle)—into a shared representation space, delineating class-specific regions. Given an input image, STHLM generates multiple query samples (blue hexagons), each mapped into this space. The nearest class embedding is identified for each sample, and the resulting frequency distribution across classes is used to assign multiple labels to the image.

**d)** Multi-label classification performance comparison between the PathGen and STHLM using varying numbers of samples (N).

**e)** Medical records consisting of image (circle)–text (triangle) pairs (connected by lines) are encoded into a joint latent space. STHLM samples blue hexagons (image queries) or blue squares (text queries) to retrieve the closest pair.

**f)** Performance benchmark evaluating the ability to retrieve correct image-text pairs, comparing the PathGen-CLIP and STHLM-PathGen across different sample counts (N).

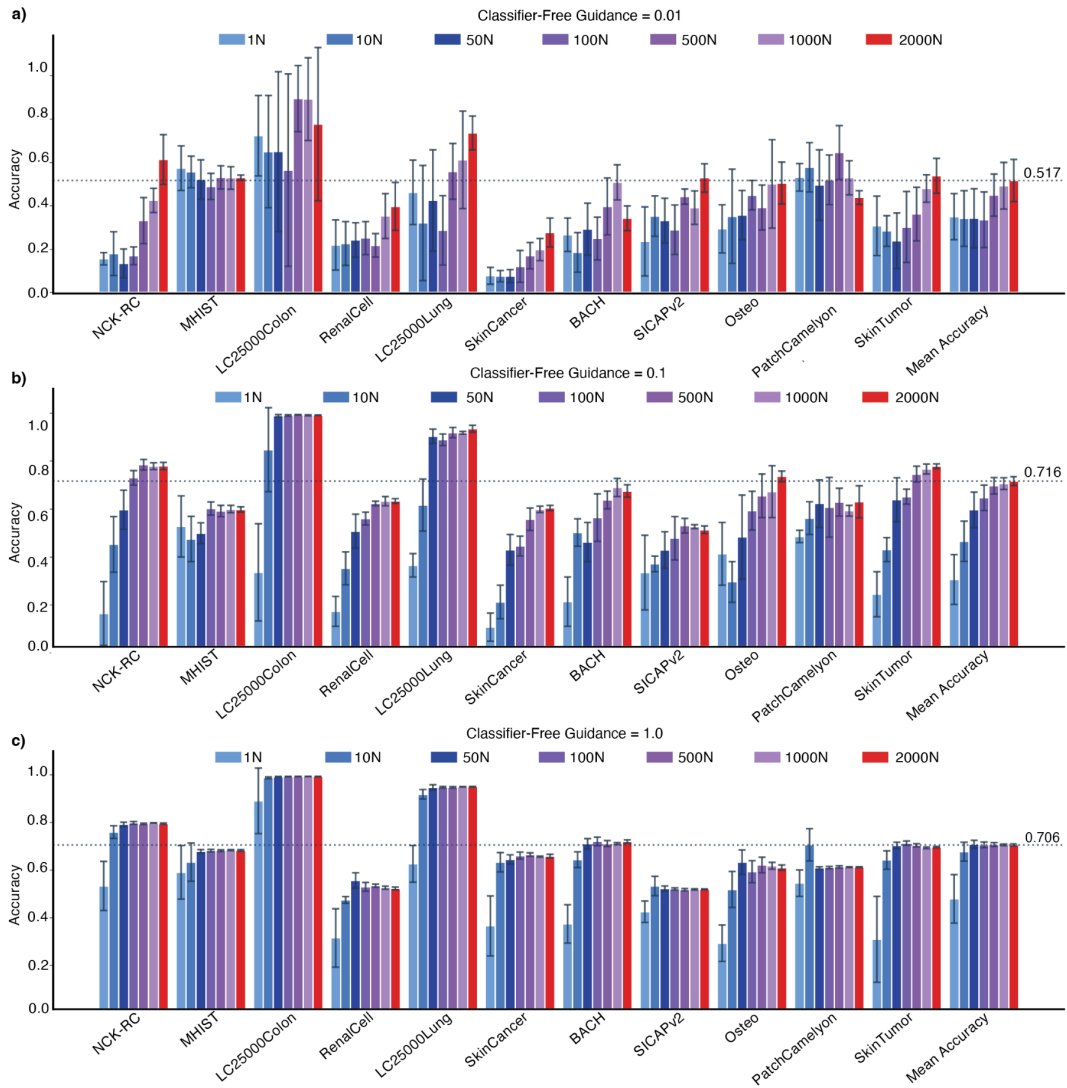

**Supplementary Figure 5: Effect of classifier-free guidance on STHLM performance across 11 histopathology benchmarks** (All bar plots display mean performance with error bars denoting ±1 standard deviation).

**a)** A guidance weight of 0.01 significantly reduces accuracy, indicating poor class fidelity.

**b)** A weight of 0.1 achieves consistently high accuracy, indicating an optimal balance between diversity and class fidelity.

**c)** A weight of 1.0 accelerates convergence but reduces accuracy relative to 0.1, suggesting that excessive class fidelity comes at the cost of diversity.

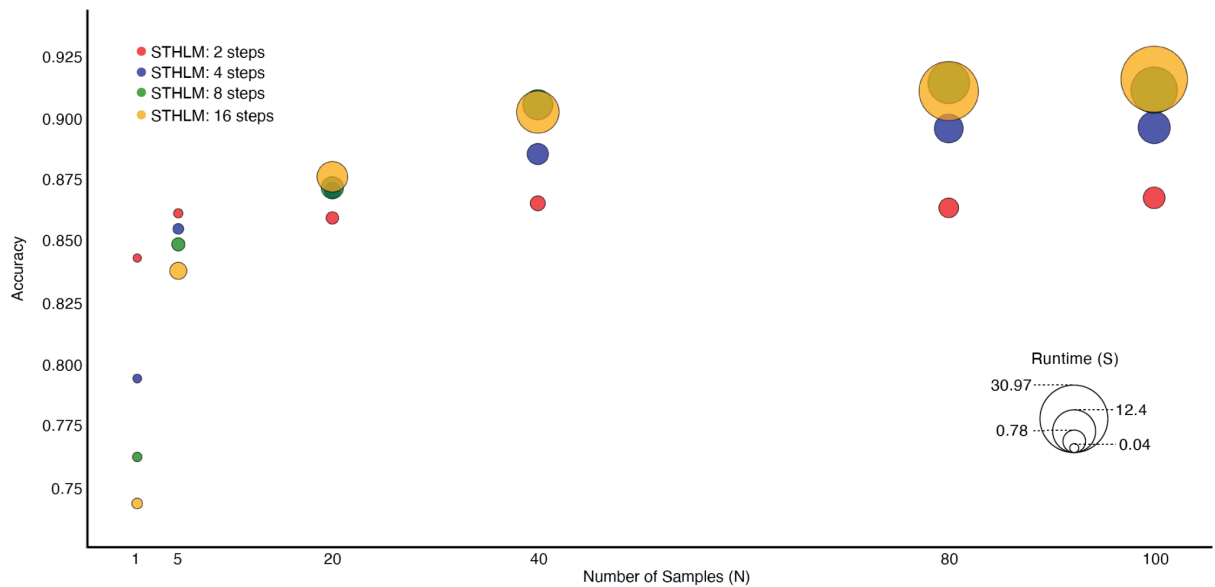

**Supplementary Figure 6: Runtime–accuracy trade-off for STHLM across varying numbers of Euler steps.** To evaluate the effect of Euler step count on classification performance, we tested STHLM on the clinical record classification task shown in Fig. 2f. The task involved 539 tissue image samples and 2,473 labeled text records (Cancer/No Cancer), which served as the reference set for labeling. We repeated the experiment five times with different random selections of the 539 tissue images and used the mean accuracy. Accuracy is plotted on the y-axis, and the number of classification samples (N) on the x-axis. We compared four discretization levels of the Euler ODE solver: 2 steps (red), 4 steps (blue), 8 steps (green), and 16 steps (yellow). Results show lower step counts yield better performance with fewer samples, likely due to increased sample diversity. In contrast, as the number of samples increases, higher step counts improve accuracy.

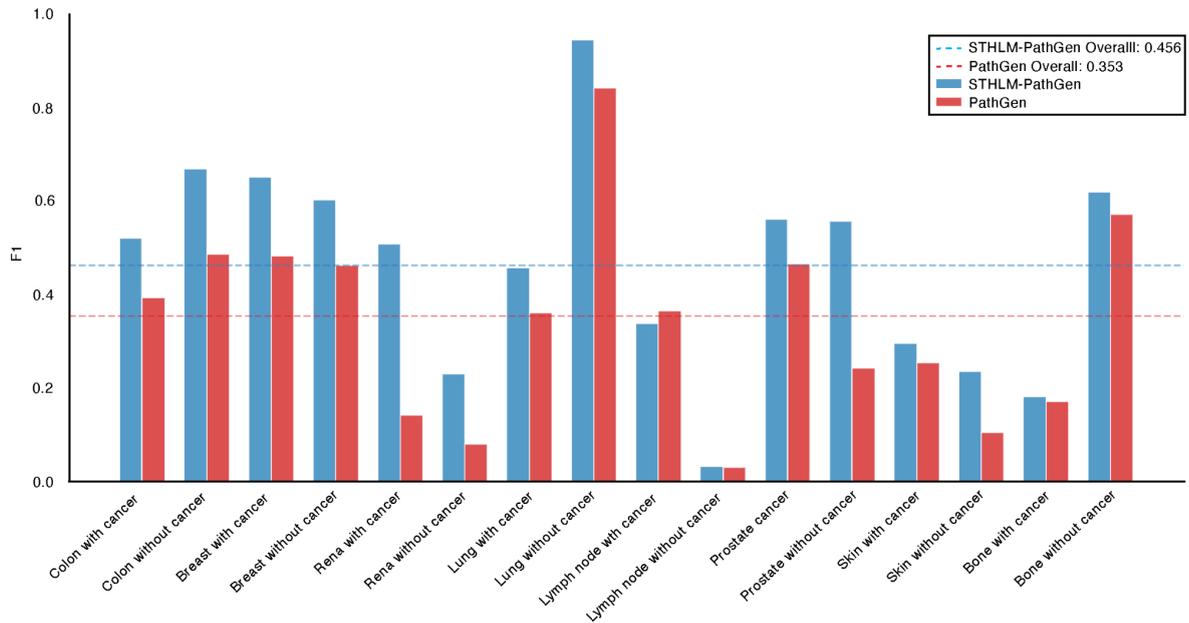

***Supplementary Figure 7: Clinical record–guided classification of tissue images.***

*Nearest-neighbor classification performance comparison between PathGen (red) and STHLM-PathGen (blue) with 40N inference samples, across eight tissue types (colon, breast, renal, lung, lymph node, prostate, skin, and bone), each further categorized as cancerous or non-cancerous, resulting in 16 classes in total. Dashed lines indicate the mean F1-score for each model.*

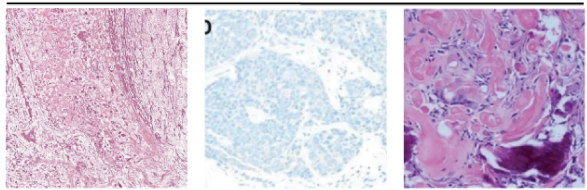

a) Query — Bone with cancer | STHLM — Bone with cancer | PathGen — Skin without cancer

All slides stained with H and E stain. An image of slide with magnification x400 demonstrate a well differentiated chondroid lesion composed of scattered, relatively bland, stromal cells/chondrocytes. Higher magnification view demonstrate angular and stellate cells set in bluish-pink chondromyxoid stroma. Note that the tumor lacks true hyaline cartilage matrix seen in enchondromas and chondrosarcomas.

(A, B). High power Slide showed masses of basaloid cells. Some of the masses are attached to the undersurface of the epithelium. These masses show sebaceous differentiation. The lobules of sebaceous glands show irregular growth pattern. Germinative epithelioid cells predominate and haphazardly arranged. There are multiple keratinous microcysts and few mitotic figures.

b) Query — Bone without cancer | STHLM — Bone without cancer | PathGen — Bone with cancer

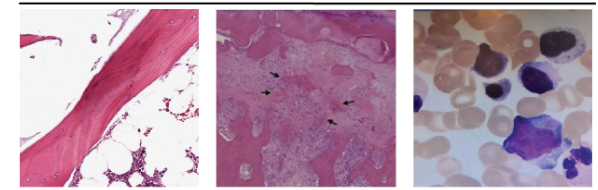

The biopsy from the proband shows a bone sample with widened trabeculae. Osteoblastic activity and irregular bone formation are seen as blue cement lines. The cement lines represent changes in the direction of new bone formation and are not pathognomonic of a specific process by themselves; they are best seen in Paget's disease but may be visible in other disorders involving rebuilding of the bone structure (Hematoxylin-Eosin stain 9.9x).

Photomicrograph of bone marrow trephine biopsy (case 1) shows hypocellular marrow spaces with 70% of fat cells. There are reduced numbers of hematopoietic cells with increased numbers of blasts (a, b).

c) Query — Breast with cancer | STHLM — Breast with cancer | PathGen — Bone with cancer

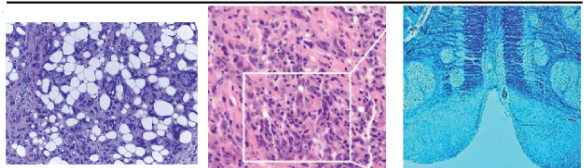

histological features of low-grade dcis from a breast biopsy showing bland homogeneous cells contained within the duct, forming rigid cell 'bridges' across the duct space in a cribriform architecture. in this case, the abnormal duct is surrounded by fibrotic stroma (hematoxylin and eosin, original magnification 100x).

H&E (100x) biopsy of the right femur lytic lesion. . The histologic section demonstrates an atypical lymphoid infiltrate comprising intermediate- to large-sized CD10-negative B lymphocytes expressing BCL6 and MUM-1, consistent with diffuse large B-cell lymphoma (DLBCL), activated B-cell type. . H&E, hematoxylin and eosin.

d) Query — Breast without cancer | STHLM — Breast without cancer | PathGen — Breast with cancer

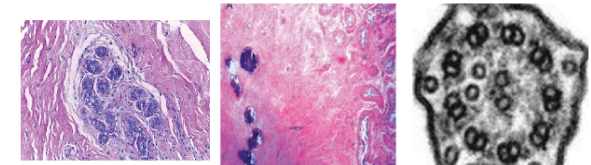

Histology of the core needle biopsy. B; Higher power (x10) shows that the intraductal proliferations consist of a monomorphic population of cohesive cells forming bridges and micropapillary structures. consistent with a diagnosis of atypical ductal hyperplasia.

Primary tumour of the breast formed by uniform cells with scant, lightly eosinophilic cytoplasm, arranged in broad gyriform trabeculae. Invasive component is visible at the top, in situ component below it (H&E, 20x).

e) Query — Colon with cancer | STHLM — Colon with cancer | PathGen — Bone with cancer

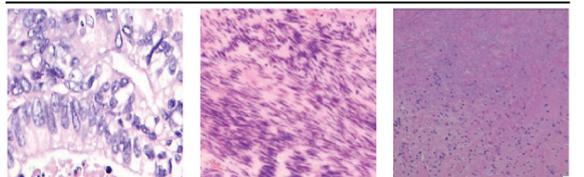

Pathological findings. (A) Low-grade mucinous tumours of the appendix have incomplete appendix muscles, and the tumour penetrates the visceral peritoneum and involves the striated muscle tissue of the abdominal wall.

Giant cell glioblastoma is composed of large, closely-arranged cells, with an eosinophilic cytoplasm and obvious nuclear atypia. There are also scattered multinucleated giant cells. Local necrosis and vascular proliferation are observed (A).

f) Query — Colon without cancer | STHLM — Colon without cancer | PathGen — Lymph node with cancer

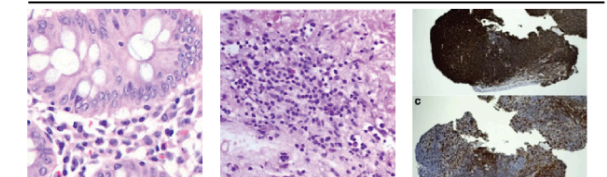

Microphotograph revealing a rectal mucosa with dense inflammatory infiltrate made of neutrophils and eosinophils. Numerous cryptic abscesses were observed (Arrow). No cytomegalovirus inclusions were observed. (HE; 100X).

(B) On higher power (100X), the atypical infiltrate consists of large lymphoid cells with prominent nucleoli and increased mitotic activity, morphologically consistent with diffuse large B cell lymphoma (DLBCL, left side). Normal small bowel villi are present on the right side for comparison.

g) Query — Renal with cancer | STHLM — Renal with cancer | PathGen — Lymph node with cancer

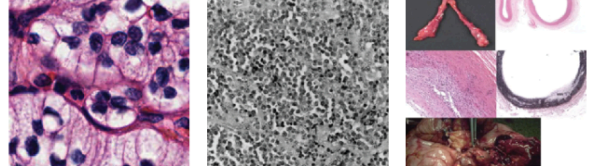

Light microscopy features of the resected pleural and renal tumors. The renal tumor was resected in another hospital 20 yr previously and diagnosed as a low-grade clear cell renal cell carcinoma. The cytoplasms of the tumor cells are D-PAS positive PAS, x 200). (D) The tumor cells of right renal mass have clear cytoplasms, distinct cell borders, and small nuclei. The cells are arranged in small solid nests (H&E, x 200).

Concomitant disseminated Langerhans cell histiocytosis and acute myeloid leukaemia. C; Tonsillar biopsy, showing infiltration by a population of atypical cells with indistinct cytoplasmic borders. High-power view (insert) showed atypical cells with nuclei containing linear grooves.

h) Query — Renal without cancer | STHLM — Renal without cancer | PathGen — Lung without cancer

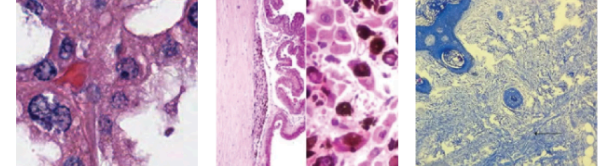

Pathology evaluation for the C1q nephropathy patient. (A) Hematoxylin-eosin staining shows increased number of cells in glomerulus. No glomerular crescent, focal segmental glomerular sclerosis, and fibrosis are noted. However, renal tubule dilation and vacuolar degeneration of epithelial cells are present, and renal interstitium is slightly infiltrated with inflammatory cells, indicating a minimal change disease (MCD).

Clinical and Diagnostic Progression in Patient with PPMS and SLE. (B-C) Subsequent to the histopathological examination of the parietal pleura, the pathology report indicates the presence of lymphocytes, histiocytes, and normal mesothelial cells, with no discernible tumor cells identified. The red arrow signifies values exceeding the normal range.

i) Query — Lung with cancer | STHLM — Lung with cancer | PathGen — Renal with cancer

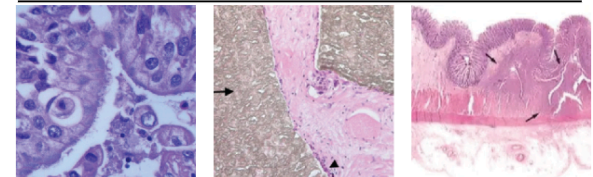

Pathological findings of biopsies obtained from the mediastinal lymph node and left upper lobe of the lung. (A) Cancer cells forming solid nests and acinar structure were observed in EBUS-TBNA specimens of the mediastinal lymph node.

Histological analysis revealed a typical morphology of a middle-grade (grade 2) clear-cell renal cell carcinoma, which confirmed the patient's tumor to be a primary neoplasm. Hematoxylin and eosin (H&E); magnification, x1,000.

j) Query — Lung without cancer | STHLM — Lung without cancer | PathGen — Lung with cancer

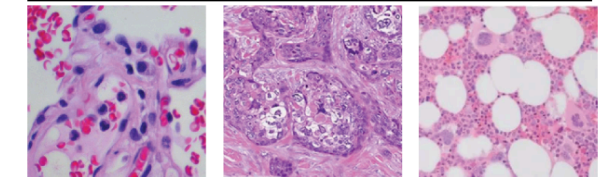

Histology of the right pleural lesion. Hematoxylin-eosin staining of the specimen revealed a marked proliferation of dilated lymphatic structures, which were consistent with the generalized lymphatic anomaly.

Pathologic response to neoadjuvant selpercatinib. Tumor specimens before and after neoadjuvant selpercatinib treatment are shown. (A) Pathologic assessment of a bronchoscopic biopsy with H&E staining before selpercatinib treatment identified a lung adenocarcinoma (x400).

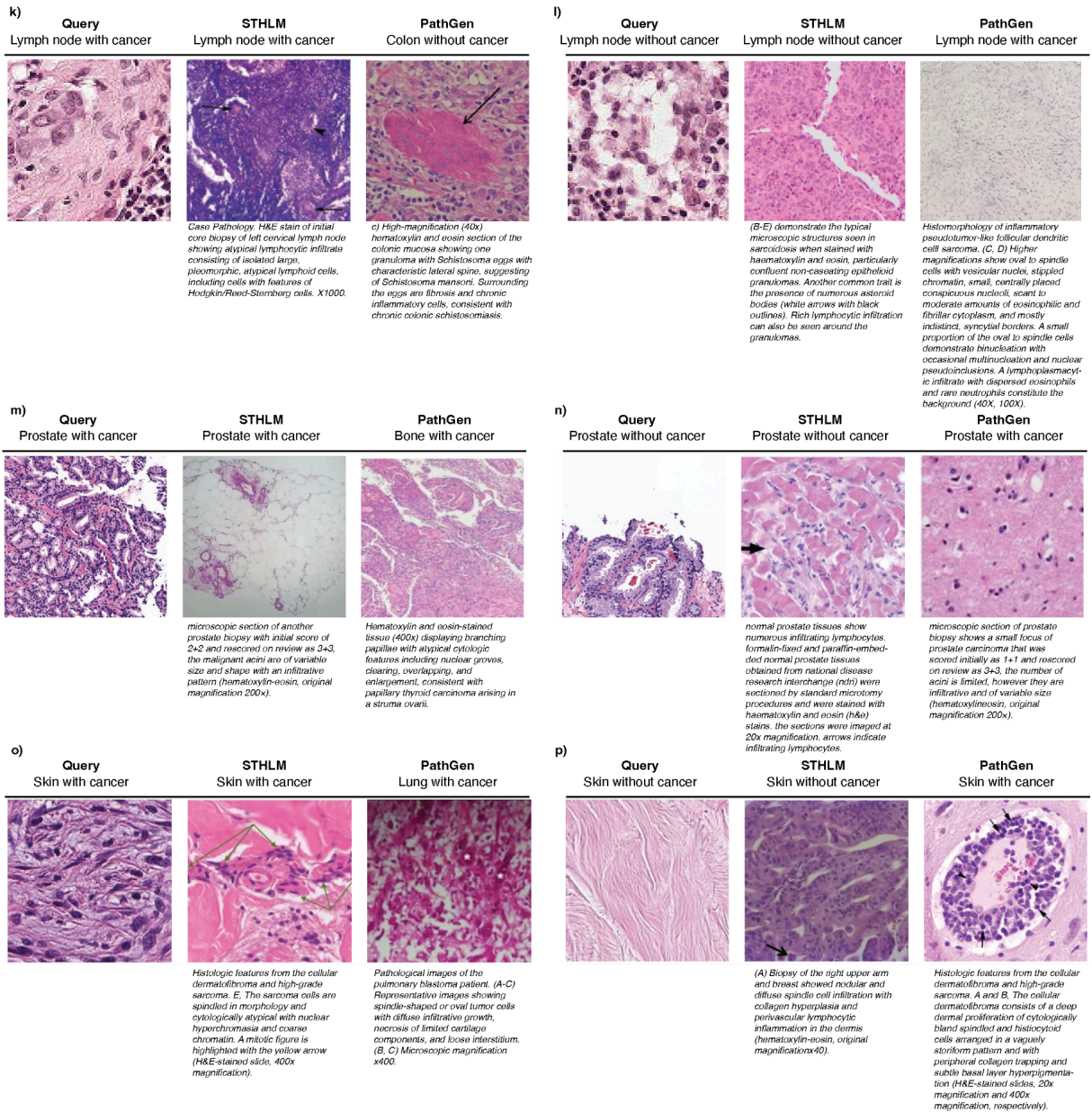

**Supplementary Figure 8: Clinical record–guided classification of tissue images where STHLM-PathGen succeeds and PathGen fails.** Examples are shown for "bone with cancer" (a), "bone without cancer" (b), "breast with cancer" (c), "breast without cancer" (d), "colon with cancer" (e), "colon without cancer" (f), "renal with cancer" (g), "renal without cancer" (h), "lung with cancer" (i), "lung without cancer" (j), "lymph node with cancer" (k), "lymph node without cancer" (l), "prostate with cancer" (m), "prostate without cancer" (n), "skin with cancer" (o), and "skin without cancer" (p).

a)
**Query**
Bone with cancer

**STHLM**
Lymph node with cancer

**PathGen**
Bone with cancer

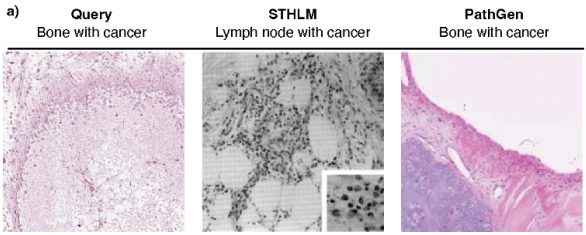

(A) Low-power pathologic slide from the second patient demonstrating the key features of colloid carcinoma, with extensive mucin pools and scattered neoplastic cells within them in the bottom of the photograph. Associated IPMN can also be seen as the ductal cells with papillary projections (arrow).

Histopathologic assessment of the right temporal lobe lesion. D; H&E stained specimen showing necrosis with viable perivascular tumor cells, more characteristic of tumor necrosis than radiation-associated necrosis (magnification, x10).

b)
**Query**
Breast with cancer

**STHLM**
Breast without cancer

**PathGen**
Breast with cancer

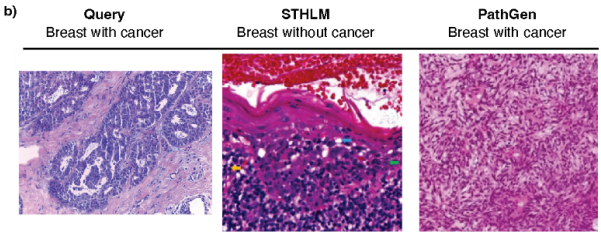

microscopic image of left breast nodule at high power, h&e 20x. the stroma of the lesion has low cellularity with no mitoses or cytologic atypia. the epithelial component shows bland cytologic features with associated myoepithelial cells.

Ductal carcinoma in situ (DCIS) within a benign phyllodes tumor. (C) Higher magnification of the DCIS demonstrates a monotonous population of epithelial cells with intermediate grade nuclei and rigid cribriform architecture. Minute calcifications are associated with DCIS (arrowhead).

c)
**Query**
Breast without cancer

**STHLM**
Skin without cancer

**PathGen**
Breast without cancer

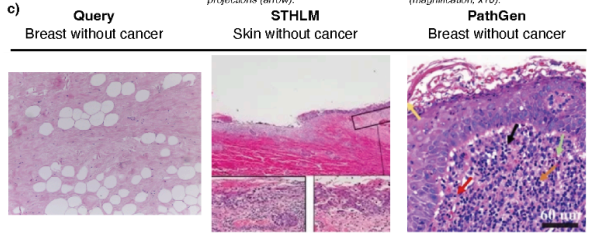

52-year-old man with silicone-induced penile sclerosing lipogranuloma. Histopathologic evaluation of the excised tissue reveals vacuoles of variable size in the dermis corresponding to exogenous substance. Multinucleated giant cells and abundant fibrosis is seen around the vacuoles (Hematoxylin and Eosin stain, magnification x20).

hematoxylin and eosin stained histology slide at 10x shows small- and medium-sized capillary-like vessels infiltrating the fat and breast extralobular parenchyma. endothelial atypia and complex anastomosing vessels are not seen. scant lymphocytes surrounding the blood vessels are present.

d)
**Query**
Colon without cancer

**STHLM**
Colon with cancer

**PathGen**
Colon without cancer

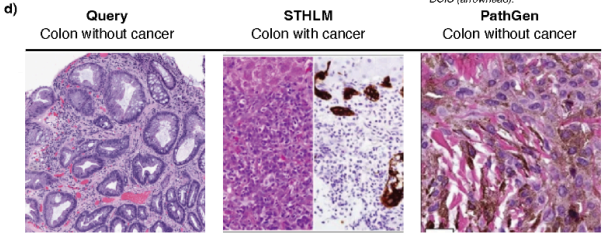

Immunohistochemistry demonstrated on vaginal biopsy. . Notes: This combination of stains is highly suggestive of a malignancy of gastrointestinal origin. (A) Moderately differentiated adenocarcinoma on H&E staining at 200x magnification.

Pathologic examination of the colon reveals ulceration with glandular distortion, widening of the distance between the crypts, and prominent infiltration of inflammatory cells, compatible with Behcet colitis (H&E, x 100).

e)
**Query**
Renal with cancer

**STHLM**
Renal without cancer

**PathGen**
Renal with cancer

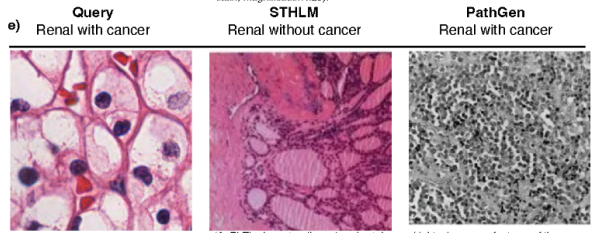

(A, B) The hematoxylin and eosin stain (H&E) of renal and bladder growth show a similar fashion of epithelial cells with clear cytoplasm and prominent nuclei arranged in nests with intervening branching vascular tissue.

Light microscopy features of the resected pleural and renal tumors. The renal tumor was resected in another hospital 20 yr previously and diagnosed as a low-grade clear cell renal cell carcinoma. The cytoplasms of the tumor cells are D-PAS positive PAS, x 200). (D) The tumor cells of right renal mass have clear cytoplasms, distinct cell borders, and small nuclei. The cells are arranged in small solid nests (H&E, x 200).

f)
**Query**
Renal without cancer

**STHLM**
Breast without cancer

**PathGen**
Renal without cancer

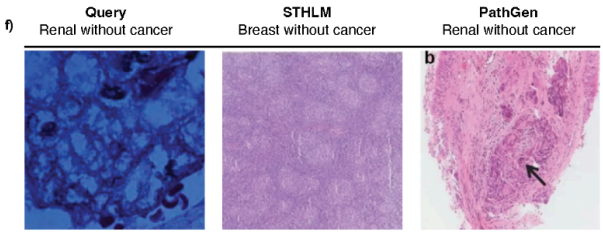

low power view showing a few granulomas with central caseous necrosis and thin peripheral rim of lymphocytes. normal breast glands showing pregnancy-associated changes can be seen at the upper edges of the photomicrograph (h&e, x80)

Histopathology seen on renal biopsy. . Light microscopy - The non-sclerotic glomeruli show nodular glomerulosclerosis. The tubular basement membranes show similar powdery granular electron-dense deposits (h).

g)
**Query**
Lung with cancer

**STHLM**
Skin with cancer

**PathGen**
Lung with cancer

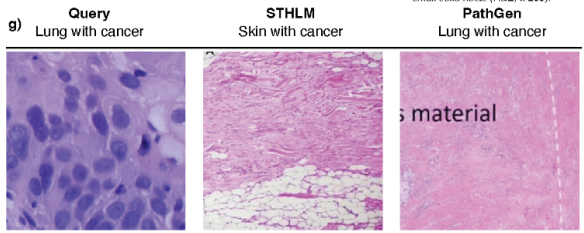

fna of right axillary skin. diff-quik stain, x60. description: cellular specimen. tridimensional clusters of large epithelial cells with prominent bizarre macronucleoli. adenocarcinoma metastatic to skin, consistent with prostate primary.

Histopathological findings of the surgical specimen. (a-c) Bronchoscopic biopsy of the left upper lobe revealing adenocarcinoma cells with mucus production. Hematoxylin and eosin staining; Original magnification 200x (a, d).

h)
**Query**
Lymph node with cancer

**STHLM**
Lymph node without cancer

**PathGen**
Lymph node with cancer

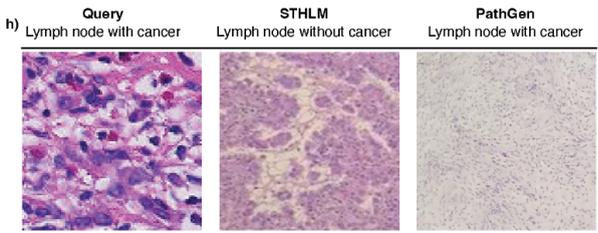

Histological features of excisional biopsy demonstrating eosinophilic infiltration. C; Same lymph node. Eosinophilic microabscess formation with hyperplasia of endothelial cells in postcapillary venules. (H&E stain, x400).

Histomorphology of inflammatory pseudotumor-like follicular dendritic cell sarcoma. (C, D) Higher magnifications show oval to spindle cells with vesicular nuclei, stippled chromatin, small, centrally placed conspicuous nucleoli, scant to moderate amounts of eosinophilic and fibrillar cytoplasm, and mostly indistinct, syncytial borders. A small proportion of the oval to spindle cells demonstrate binucleation with occasional multinucleation and nuclear pseudoinclusions. A lymphoplasmacytic infiltrate with dispersed eosinophils and rare neutrophils constitute the background (40X, 100X).

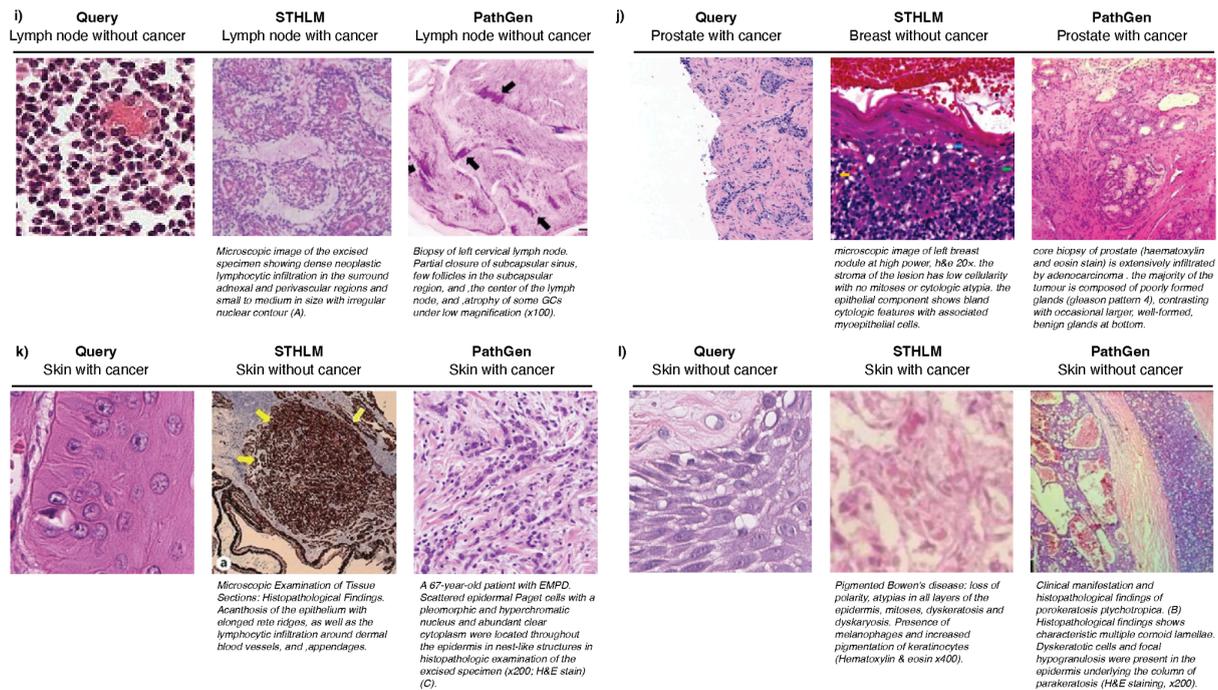

**Supplementary Figure 9: Clinical record–guided classification of tissue images where PathGen succeeds and STHLM-PathGen fails.** Examples are shown for "bone with cancer" (a), "breast with cancer" (b), "breast without cancer" (c), "colon without cancer" (d), "renal with cancer" (e), "renal without cancer" (f), "lung with cancer" (g), "lymph node with cancer" (h), "lymph node without cancer" (i), "prostate with cancer" (j), "skin with cancer" (k), and "skin without cancer" (l). No examples were found for "bone without cancer", "colon with cancer", "lung without cancer", or "prostate without cancer" where PathGen succeeded and STHLM-PathGen failed.